\documentclass[preprint,showpacs,amsmath,amssymb,aps]{revtex4-1}

{}
\usepackage{amsmath}
\usepackage{graphicx}
\usepackage{dcolumn}
\usepackage{bm}
\usepackage[utf8]{inputenc}
\usepackage{listings}
\usepackage{subcaption}
\usepackage{color}
\usepackage[normalem]{ulem}

\usepackage{setspace}
\begin{document}
\title{Fermi-L\"owdin orbital self-interaction correction using the strongly 
constrained and appropriately normed meta-GGA functional}

\author{Yoh Yamamoto$^*$}
\author{Carlos M. Diaz$^{*\S}$}
\author{Luis Basurto$^*$}
\author{Koblar A. Jackson$^\dagger$}
\author{Tunna Baruah$^{*\S}$}
\author{Rajendra R. Zope$^{*\S{a)}}$}
\affiliation{$^*$Department of Physics, The University of Texas at El Paso, El Paso, Texas, 79968}
\affiliation{$^{\S}$Computational Science Program, The University of Texas at El Paso, El Paso, Texas, 79968}
\affiliation{$^\dagger$Physics Department and Science of Advanced Materials Program, Central Michigan University, Mt. Pleasant Michigan, 48859}
\email{$^{a)}$rzope@utep.edu}

\begin{abstract}
	Despite the success of density functional approximations (DFAs) in describing the electronic properties
	of many-electron systems, the most widely used approximations suffer from 
	self-interaction errors (SIE) that limit their predictive power. %We have recently implemented 
	Here we describe the effects of removing SIE from  the strongly constrained and appropriately normed (SCAN)  meta-generalized gradient approximation (GGA) using
	%that satisfies all 17 known exact constraints and is appropriately normed 
       the Fermi-L\"owdin Orbital Self-Interaction Correction (FLOSIC) 
	method. FLOSIC is a size-extensive implementation of the Perdew-Zunger self-interaction 
	correction (PZ-SIC) formalism.  We find that FLOSIC-SCAN calculations require careful 
	treatment of numerical details and describe an integration grid that yields reliable accuracy with this approach. We investigate the performance of FLOSIC-SCAN for predicting a wide array of properties %including total atomic energies, ionization potentials, electron affinities, reaction energies and atomization energies of molecules. 
	and find that it provides better results than FLOSIC-LDA and FLOSIC-PBE in nearly all cases.  It also gives better predictions than SCAN for orbital energies and dissociation energies where self-interaction effects are known to be important, but total energies and atomization energies are made worse. 
        %; however, improved description of the asymptotic region results
	%in highest occupied orbitals eigenvalues that are closer to the experimental first ionization energies.
	For these properties, we also investigate the use of the self-consistent FLOSIC-SCAN density in the SCAN functional and find that this DFA@FLOSIC-DFA approach yields improved 
	%atomization energies, reaction energies, ionization potentials and electron affinities for the semilocal PBE and SCAN functionals 
	results compared to pure, self-consistent SCAN calculations. %Similar improvements in the total atomic energies, atomization energies, electron
%	affinities and reaction enegies over parent SCAN functional were observed when self-consistent FLOSIC-SCAN density
	%is used in the uncorrected SCAN functional.
	%\textcolor{red}{ Thus DFA@FLOSIC-DFA procedure is a pragmatic approach to obtain good 
	%energetics from the FLOSIC-DFA calculations particularly for the properties that are made worse 
	% by application of SIC.}
	  Thus FLOSIC-SCAN provides improved results over the parent SCAN functional in cases where SIEs are dominant, and even when they are not, if the SCAN@FLOSIC-SCAN method is used.
	  %equilibrium properties are obtained using the 
	 %approach while employing FLOSIC-SCAN for cases where SIE is dominant.}
	%Interestingly, using the self-consistent
	%FLOSIC-SCAN density in the uncorrected SCAN functional  
	%provides improved results for total energies and  atomization energies over standard SCAN, as judged by  mean absolute errors.

	%The later are already available in the self-consistent
	%FLOSIC calculations. 
	%Such a procedure preserves the correct $-1/r$ asymptotic nature of the potential 
	%while maintaining the accuracy of the total energies and atomization energies of the parent 
	%functional. 
       % For the SCAN functional this procedure provides a slight improvement in the total and atomization energies over the  
       %	standard SCAN functional.

\end{abstract}
\date{\today}
%\pacs{To be determined}

%\pdfinfo{%
%  /Title    ()
%  /Author   ()
%  /Creator  ()
%  /Producer ()
%  /Subject  ()
%  /Keywords ()
%}

%\begin{document}
\maketitle

\section{Introduction}

%\section{Introduction}

%*DFT has self-interaction errors.
Density functional theory (DFT) has been widely used to study the electronic structure of various types of materials from atoms and molecules to 
nanostructures to periodic materials. 
The popularity of DFT stems from its low computational expense combined with relatively good accuracy. 
The self-interaction error (SIE) that arises from  the density functional approximations (DFAs) of the exchange-correlation 
functional is well-documented \cite{PhysRevB.23.5048}. This error arises since the self-Coulomb energy is not completely canceled by 
the self-exchange energy when the exact, but unknown, exchange-correlation functional is approximated.
This leads to a number of problems. For example, the one-electron potential in DFA does not have the correct asymptotic behavior due to the presence of the SIE, leaving the highest occupied orbitals in stable anions unbound as a result.

% however, it has been known that DFT contains self-interaction errors (SIE). According to Pauli exclusion principle, Coulomb and exchange self-interactions should sum up to zero. DFT contains SIE because of an approximation in exchange functional.
%SIE cause several problems in DFT. 
%The Khon-Sham potential does not have a correct long-range behavior with SIE.
%SIE is also known to cause delocalization errors in DFT calculations.  
%One would expect SIE also affects the accuracy of electron density and band gaps.

%*Perdew-Zunger introduced self-interaction correction (SIC).
%There have been methods to address self-interaction correction (SIC) in DFT. %as early as 1930 where \cite{Fock1930}.
The Perdew-Zunger self-interaction correction formalism (PZ-SIC) is a one-electron self-interaction-free 
approximation where an orbital by orbital correction is applied to the DFA total energy \cite{PhysRevB.23.5048}. A  number of implementations of SIC to DFT
exist \cite{doi:10.1063/1.481421, doi:10.1063/1.1327269, doi:10.1063/1.1370527, doi:10.1063/1.1468640, doi:10.1021/jp014184v,PhysRevA.55.1765,doi:10.1080/00268970110111788, Polo2003, doi:10.1063/1.1630017, B311840A,doi:10.1063/1.1794633, doi:10.1063/1.1897378, doi:10.1063/1.2176608, doi:10.1063/1.2204599,doi:10.1002/jcc.10279,PhysRevA.45.101, PhysRevA.46.5453, PhysRevA.47.165}, including a 
recent implementation by J\'onsson \textit{et al.} using complex orbitals that has shown promising results \cite{doi:10.1021/acs.jctc.6b00347}.
%An orbital-dependent single-particle potential is introduced, and the self-Coulomb energy is canceled in an orbital-by-orbital manner.
The PZ-SIC formalism corrects SIE, but it also leads to an orbital-dependent theory since the orbital-dependent total energy is not invariant under a unitary transformation of the 
occupied %Kohn-Sham 
orbitals.  The set of orbitals that yields the minimum self-interaction corrected total energy therefore must be found.  Pederson \textit{et al.} showed that these minimum-energy  local orbitals  
  satisfy additional pairwise conditions known as the localization equations (LE) \cite{doi:10.1063/1.446959, doi:10.1063/1.448266}.
Varying the N$^2$ elements of a unitary transformation to find local orbitals that satisfy the LE is a process that scales poorly with increasing numbers of orbitals, making the solution of 
the LE computationally challenging.  Another problem with traditional PZ-SIC is that it is not formally size-extensive.
%The SI corrected orbitals are 
%\textcolor{red}{orthonormal}.
%{\bf more} localized and are also not orthogonal.
%
The canonical Kohn-Sham (KS) orbitals tend to delocalize with increasing system size.  In the limit of very large sizes and very delocalized orbitals, the correction terms in PZ-SIC tend to zero\cite{Perdew1990113}.  This leads to a breakdown of size extensivity when the lowest-energy correction for a single atom is positive. 
An alternative approach to solving the LE in PZ-SIC was introduced by Pederson, Perdew, and Ruzsinszky through the use of Fermi-L\"owdin orbitals (FLO)
\cite{doi:10.1063/1.4869581} to evaluate the PZ-SIC total energy.  (The resulting method is known as FLOSIC.)
The FLOs are orthonormal local orbitals that are a linear combination of Fermi orbitals (FO). The FOs 
depend on the density matrix and spin density at certain points in space called Fermi orbital descriptors (FODs). 
The FOs are obtained from the KS orbitals as
\begin{equation}\label{eq:fod}
 \phi_{i\sigma}^{FO}(r) =\frac{\sum_j^{N_\sigma}\psi_{j\sigma}^* ({\bf a}_{i\sigma}) \psi_{j\sigma}(\vec{r})}{\sqrt{\rho_\sigma({\bf a}_{i\sigma})}}
\end{equation}
where $\psi_{j\sigma}$, $\rho_\sigma$, $\bf{a}_{i\sigma}$, $N_\sigma$  denote KS orbital, 
total electron density, FOD, and number of occupied orbitals of spin $\sigma$, respectively. The FO transformation is unitarily invariant, \textit{i.e.} the same set of FO's is produced by any orthonormal set of orbitals spanning the occupied space. The total energy in FLOSIC therefore depends on the FOD positions and the LE do not need to be applied.  In addition, because the FLOs are localized, the FLOSIC method restores size extensivity \cite{doi:10.1063/1.4869581}.
%
%
%Since this set of local orbitals depends
%on quantities that are invariant under unitary transformation, the total energy %not the SIC hamiltonian
%remains unitarily invariant. The use of FLOs has the advantage that localization equations do not need to be applied. 

The FO are determined by the positions of
the FODs; therefore, only 3N variables are needed to determine 
the optimal set of local orbitals, compared to N$^2$ coefficients of a unitary transformation needed in traditional PZ-SIC.  Thus, in principle, FLOSIC provides a computationally simpler way to incorporate the self-interaction correction.
%Since the FLO implementation of SIC (FLOSIC) 
%incorporates the local orbitals that are invariant under a unitary transformation and i
%bypasses the need to solve the LEs, in principle, it provides a computationally simpler way to incorporate the self-interaction correction.  
In  practical FLOSIC calculations, optimal FOD positions are found using gradients of the energy with respect to FOD positions, in a procedure analogous to molecular geometry optimizations 
\cite{doi:10.1063/1.4907592,PEDERSON2015153}. A number of studies have been conducted using the FLOSIC method \cite{doi:10.1063/1.4869581,doi:10.1063/1.4996498,magnetochemistry3040031,doi:10.1080/00268976.2016.1225992,doi:10.1063/1.4947042,doi:10.1021/acs.jpca.8b09940, doi:10.1063/1.5050809, PhysRevA.100.012505,doi:10.1002/jcc.26008}.

To date, FLOSIC has been applied mostly to the LDA level of theory where nearly all properties of atoms and molecules are significantly improved %are found for atomic total energies and atomization energies of molecules 
\cite{doi:10.1063/1.4996498,PhysRevA.95.052505,doi:10.1063/1.4907592}. 
On the other hand, SIC-based improvements are known to be less uniform with semilocal generalized gradient approximations (GGA) and  meta-GGAs \cite{PhysRevA.84.050501,doi:10.1063/1.1794633,doi:10.1063/1.5087065}.
Recently, Perdew and coworkers have provided insight into this problem \cite{doi:10.1063/1.5087065}, %of using SIC in conjunction with semilocal functionals by suggesting 
showing that the lobed one electron densities needed for applying SIC are problematic for semilocal functionals such as the Perdew, Burke, and Ernzerhof (PBE) \cite{PhysRevLett.77.3865,PhysRevLett.78.1396} GGA and the strongly constrained and appropriately normed (SCAN) \cite{PhysRevB.54.16533, *PhysRevB.57.14999, PhysRevLett.115.036402} meta-GGA.
   While the use of complex orbitals can lessen the problem, it does not 
   eliminate it \cite{doi:10.1063/1.5087065}.  In related work, Santra and Perdew showed applying SIC to a semilocal functional causes appropriate norms that are built in to the functional to be violated \cite{doi:10.1063/1.5090534}.  

%   This explains why the semilocal functionals perform poorly with SIC \textcolor{red}{for certain electronic properties although they improve the other areas}. 

%Kl\"{u}pfel \textit{et al.} studied atoms using PZ-SIC and PBE \cite{PhysRevA.84.050501}. They showed that orbital-dependent density functionals %require 
%\textcolor{red}{with inclusion of complex local orbitals find the lower energy  in the variational minimization process
%than only when the real local orbitals are used}.
%
%In this work, we study the performance of the FLOSIC method on the {\em electron density} with LSDA, PBE, and SCAN functionals through 
%\textcolor{red}{%We perform 
%an indirect comparison of densities by comparing the performance of DFT functionals for calculations of several physical quantities using the self-consistent, 
%self-interaction corrected electron densities. }
%
%
%{\textcolor{blue}{
% 
%On the other hand, the SCAN functional is known to satisfy all the known constraints. In that view, it is instructive to see how the self-interaction correction changes the electronic energies as well as densities
%in conjunction with a physically accurate functional. 
%}}
%
%
Because these recent developments may lead to new approaches to implementing SIC and because SCAN is the most successful nonempirical 
semilocal functional for predicting the properties of atoms, molecules, 
and solids, it is important to thoroughly benchmark the performance of SCAN when used with the existing FLOSIC methodology. We note that although some initial applications of FLOSIC-SCAN were included in the recent publications \cite{doi:10.1063/1.5087065,doi:10.1063/1.5090534}, this article presents the details of the FLOSIC-SCAN implementation for the first time, including a description of refinements to the numerical integration grid that are necessary to insure accurate results. It also gives  %know how self-interaction correction affects the results obtained using SCAN.
%List what has been done for SIC-SCAN below
%Recently, some initial results for SCAN with SIC for atomization energies and barrier heights appeared \cite{doi:10.1063/1.5087065}. 
%There is also a report on enthalpies of formation and HOMO eigenvalues using SIC-SCAN \cite{2019arXiv190502631S}. 
a full account of how FLOSIC-SCAN performs for a number of properties such as atomic energies, $\Delta$-SCF ionization potentials and electron affinities, ionization potential estimates from the HOMO energies of atoms and molecules, dissociation energies using benchmark sets that are known to be sensitive to SIEs, and atomization energies. In all cases, we compare the performance of FLOSIC-SCAN to that of FLOSIC-LDA and FLOSIC-PBE and the uncorrected SCAN functional.
%This work reports the details of implementation  and the performance assessment of FLOSIC-SCAN for a number of properties such as atomic energies, $\Delta$-SCF ionization potentials and electron affinities, ionization potential estimates from the HOMO energies of atoms and molecules, reaction energies and atomization energies. It also reports the first results of SCAN calculations using the self-consistent self-interaction corrected density. The goal is to provide a complete picture of how FLOSIC-SCAN performs with respect to SCAN and FLOSIC with other functionals.
%It emphasizes the need for refinements to the integration grid that are necessary to ensure accurate results. 
%\textcolor{black}{
%The need of dense numerical grid in the FLOSIC-SCAN
%calculations makes it  
%more expensive compared to the FLOSIC-LSDA, 
We  also examine the effectiveness of using FODs
%of using the Fermi-L\"owdin orbital descriptors 
optimized at the FLOSIC-LDA level 
in FLOSIC-SCAN calculations.

Finally, we also investigate the quality of the self-consistent FLOSIC-DFA electron density
by using it in place of the corresponding self-consistent DFA density in the parent DFA functional.  Since the FLOSIC method restores the correct asymptotic behavior to the DFA potential for a localized system, it is expected to improve the quality of the density in the asymptotic region.  Hence, the more physically correct electron density from FLOSIC,
   %Since the FLOSIC method 
   %can give electron density with accurate asymptotic behavior, 
   when combined with an accurate
   functional such as SCAN, may lead to improved estimates of total energies
   by removing density driven errors \cite{VERMA201210, PhysRevLett.111.073003}.  Our results show that using the FLOSIC density in the parent functional often leads to electronic properties near equilibrium that are improved over those of the parent functional. 
 %\textcolor{blue}{ 
% The errors in density functional theory calculations can arise from approximations to the exchange-correlation functional and also from the 
%approximate electron density produced by the DFAs. %}
%   Verma \textit{et al.} have found success in reducing density driven errors by 
%   using accurate electron densities \cite{VERMA201210, PhysRevLett.111.073003}.  
%   Since the FLOSIC method restores the correct asymptotic behavior to the DFA potential for a localized system, it is expected to improve the quality of the density in the asymptotic region.
%   Hence, the more physically correct electron density from FLOSIC,
   %Since the FLOSIC method 
   %can give electron density with accurate asymptotic behavior, 
%   when combined with an accurate
%   functional such as SCAN, may lead to good estimates of total energies
%   by removing density driven errors.
%
%In this work, we study the performance of the FLOSIC method on the {\em electron density} with LSDA, PBE, and SCAN functionals indirectly  
%We perform 
% by comparing the performance of DFT functionals on calculations of several physical quantities using the self-consistent, 
%self-interaction corrected electron densities.  %We note that such a procedure is similar in spirit of correcting/scaling down the SIC approaches
%reported in literature\cite{doi:10.1063/1.2176608, doi:10.1063/1.4752229}.

This article is organized as follows.
In Sec. \ref{cpsetup}, we present our computational method and also discuss the
%\textcolor{red}{The} 
 implementation of SCAN in the FLOSIC code.
% and how the issue can be resolved.}
Calculated data for the atoms and their ionization potentials and electron affinities using the FLOSIC method are discussed 
in Sec. \ref{atomandip}.
FLOSIC total and atomization energies of selected molecules are presented and discussed in
 Sec. \ref{atomizationenergymol}.
%\textcolor{red}{An accurate numerical grid discussed in Sec. \ref{cpsetup} is used to obtain all the FLOSIC-SCAN results.}
FLOSIC dissociation energies are presented in Sec. \ref{sec:sie11}.
Finally in Sec. \ref{evaluehoorbitals}, we discuss the eigenvalues of the highest occupied molecular orbitals  using FLOSIC.

%\section{Computational Method}\label{cpsetup}

\section{Computational Method}\label{cpsetup}

%*Code: FLOSIC version 0.1.2.
All of the results presented in this manuscript are calculated with the FLOSIC code, which is based on the UTEP version of the NRLMOL code \cite{FLOSICcode}, a Gaussian orbital-based electronic structure code \cite{PhysRevB.41.7453, PhysRevB.42.3276, doi:10.1002/(SICI)1521-3951(200001)217:1<197::AID-PSSB197>3.0.CO;2-B}.  Among the features included in this version is an interface to the exchange-correlation library called LIBXC. The latter provides access to a large number exchange-correlation functionals \cite{LEHTOLA20181, MARQUES20122272}. 
The FLOSIC code inherits the optimized Gaussian basis sets of NRLMOL \cite{PhysRevA.60.2840} and an accurate numerical integration grid scheme \cite{PhysRevB.41.7453}.  
In all of our calculations, the default NRLMOL basis sets are used. 
   A recent study  which studied ionization potentials and enthalpies of formation using FLOSIC approach, the default NRLMOL basis 
   set was found to provide results comparable to the cc-pVQZ basis set \cite{doi:10.1002/jcc.25586}.
The SIC calculations require finer mesh as orbital densities are involved in calculation of orbital depedent 
potentials. A default NRLMOL mesh for FLOSIC calculation, on average, has 25000 grid points per atom.  This results in integration of charge density that is accurate to the order of $10^{-8} e$.
The exchange-correlation (XC) functionals used in this study are the LSDA implementation of Perdew and Wang (LDA) \cite{PhysRevB.45.13244}, Perdew, Burke \& Ernzerhof (PBE) \cite{PhysRevLett.77.3865, PhysRevLett.78.1396}, and SCAN \cite{PhysRevLett.115.036402}. 

\begin{figure}[!htb]
\centering
\includegraphics[width=\columnwidth]{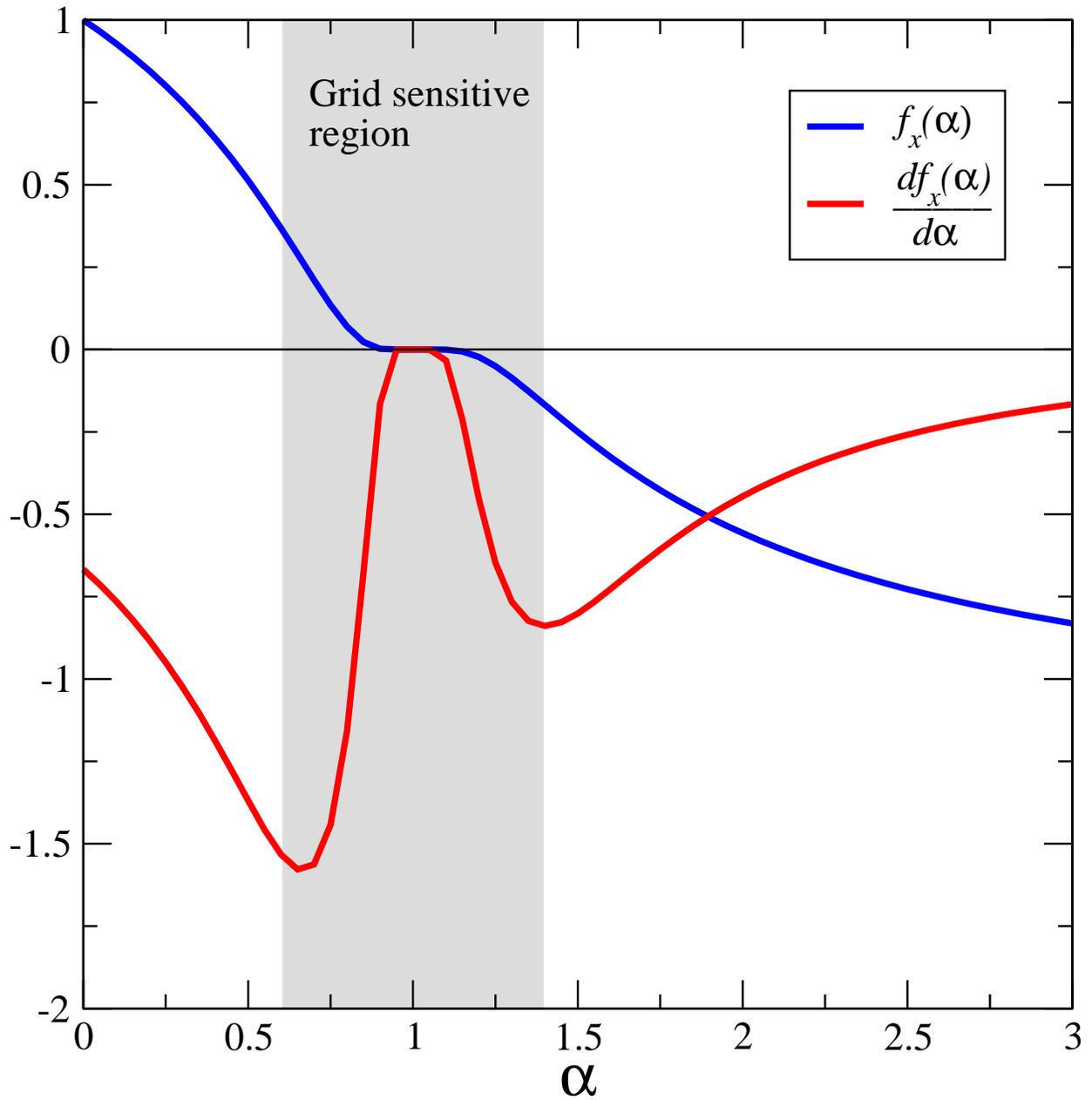}
\caption{A plot of $f_x (\alpha)$ (Eq. \ref{eqn_fx}) and ${df_x(\alpha)}/{d\alpha}$ used in the SCAN exchange enhancement factor. A large oscillation of ${df_x(\alpha)}/{d\alpha}$ is seen near $\alpha=1$. \label{scanmesh}}
%\vspace{20mm}
\end{figure}

FLOSIC calculations require an initial set of trial FOD positions. Whenever they are available, previously reported FOD positions are used as starting points. In other cases, FODs are generated from scratch and further optimized using a conjugate gradient algorithm. We use the convergence criteria of $10^{-6}$ Ha on the FLOSIC total energy for these optimizations.  We find that FLOSIC-LDA optimized FOD positions are typically a good starting point for FLOSIC-PBE and FLOSIC-SCAN calculations. For example, the FOD positions for neutral atoms shifted an average of only 0.073 Bohr  after  optimization with FLOSIC-SCAN, while keeping similar overall arrangements. 
Meta-GGA functionals, including SCAN, are sensitive to the numerical details of a calculation, and this sensitivity extends to FLOSIC-SCAN calculations. 
The standard variational integration mesh method \cite{PhysRevB.41.7453} employed in the FLOSIC code provides good accuracy for the LSDA and PBE functionals, but not  for SCAN calculations. 
Semilocal meta-GGA functionals use 
a dimensionless variable defined as
\begin{equation}
 \alpha = \frac {\tau - \tau^W}{\tau^\text{unif}} > 0
\end{equation}
where $\tau$ is the kinetic energy density, %and $\tau^W$ and $\tau^\text{unif}$ are Weizs\"acker and uniform electron gas kinetic energy densities respectively.
$\tau^W=|\vec{\nabla} \rho|^2/8\rho$ is the Weizs\"acker kinetic energy density, and $\tau^\text{unif}=(3/10)(3\pi^2)^{2/3} \rho^{5/3}$ is the kinetic energy density at the uniform-density limit.
The numerical challenges of using SCAN are related to changes in $\alpha$. 
Recently, Bart\'ok and Yates showed that the numerical instabilities arising from switching function in SCAN can be eliminated by modifying the switching function \cite{doi:10.1063/1.5094646}; however, such modification results in violation of some exact constraints.
The exchange enhancement factor of SCAN has a mathematical form given as
\begin{equation}
F_x(s,\alpha)= \{ h_x^1(s,\alpha)+ f_x(\alpha) [ h_x^0 - h_x^1(s,\alpha) ] \} g_x(s), 
\end{equation}
\begin{equation}\label{eqn_fx}
f_x(\alpha) = \exp \left[- \frac{c_{1x} \alpha}{ 1-\alpha}\right] \theta(1-\alpha) -d_x \exp \left[\frac{c_{2x}}{1-\alpha} \right] \theta(\alpha - 1),
\end{equation}
\begin{equation}
s=\frac{|\vec{\nabla} \rho|}{2(3\pi^2)^{1/3}\rho^{4/3}}
\end{equation}
where $h_x^1 (s,\alpha)$ is a function of $s$ and $\alpha$, $g_x(s)$ is a function of $s$, $h_x^0 = 1.174$, $s$ is dimensionless density gradient,
$c_{1x}$, $c_{2x}$, $d_x$ are interpolation parameters, and $\theta(x)$ is a step function of $x$ \cite{PhysRevLett.115.036402}. Figure \ref{scanmesh} shows $f_x(\alpha)$ (Eq. (\ref{eqn_fx})) and its derivative, $\frac {df_x(\alpha)}{d\alpha}$, as functions of $\alpha$. A large oscillation of $\frac {df_x(\alpha)}{d\alpha}$ is seen near $\alpha = 1$.
A high density of grid points is needed in the areas where the $\frac {df_x(\alpha)}{d\alpha}$ term changes rapidly in space, and similarly for the $\frac {df_c(\alpha)}{d\alpha}$ function used in the correlation term.
%
%In general, a large number of grid points are needed for areas where alpha is close to 1. 
The enhanced mesh used in the FLOSIC code was designed to provide this. 
To obtain numerically converged results, following procedure was adopted. We begin by adding radial points with uniform increments until the integrals are converged.
This is a brute force approach of mesh generation.
This is done to eliminate any assumption about the problematic ($\alpha \approx 1$) region. We then decrease the number of radial grid points in the region farther from the nuclei by maintaining the same grid density in the problematic ($\alpha \approx 1$) region. It is ensured that the integrals accuracy remains 
same ($10^{-8}$ Ha for exchange-correlation energy) while
reducing the grid density. This approach has worked well but still results in a numerical mesh that is approximately three to six times larger than the default variational mesh. 
The SCAN mesh used in this work is roughly 140000 grid points per atom. This results in integration of charge density which is accurate in the order of $10^{-10} e$.
Further improvement of
 the numerical grid to reduce the need of such dense grid is being explored and will be reported in future.
%There is scope to further improve this approach. We will explore this in future.}
%
%The mesh used in SCAN calculations is approximately three times larger than the default variational mesh so that all the sensitive regions are correctly described. This is sufficient to integrate $E_{XC}$ on the grid to an  accuracy of $10^{-8}$ Ha.
%

\subsection{Meta-GGA implementation}

The meta-GGA exchange-correlation energy has the form given as 
\begin{equation}\label{ExcMGGA}
% E_{XC}[\rho] =  \int e_{XC}(\rho(r),\nabla\rho(r),\tau(r))dr ,
 E_{XC}[\rho_\uparrow,\rho_\downarrow] =  \int e_{XC}(\rho_\uparrow, \rho_\downarrow, \vec{\nabla}\rho_\uparrow, \vec{\nabla}\rho_\downarrow ,\tau_\uparrow, \tau_\downarrow)d \vec{r} 
\end{equation}
where $e_{XC}$ is the exchange-correlation energy density function, $\rho_\uparrow$ and $\rho_\downarrow$ are electron spin densities, and $\tau_\uparrow$ and $\tau_\downarrow$ are kinetic energy density. The kinetic energy density is calculated from the KS orbitals $\psi_i$ as
\begin{equation}
 \tau(\vec{r}) = \frac 1 2 \sum_i \vec{\nabla} \psi_i (\vec{r}) \cdot  \vec{\nabla} \psi_i (\vec{r}).
\end{equation}
%where $\psi$ is Kohn-Sham (KS) orbital.
%In semilocal meta-GGA functional, $\tau$ is used in a dimensionless variable $\alpha$ defined as
%\begin{equation}
% \alpha = \frac {\tau - \tau^W}{\tau^\text{unif}} > 0
%\end{equation}
%where $\tau^W=|\nabla \rho|^2/8\rho$ is the Weizs\"acker kinetic energy density, and $\tau^\text{unif}=(3/10)(3\pi^2)^{2/3} \rho^{5/3}$ is the kinetic energy density at the uniform-density limit.
To obtain the exchange-correlation potential, functional derivatives of $E_{XC}$ are required.
In the case of Eq. (\ref{ExcMGGA}), the functional derivative of exchange-correlation energy with respect to density is
\begin{equation}\label{dExcdrho}
\begin{aligned}
 \frac {\delta E_{XC}[\rho]} {\delta\rho(\vec{r})} =& \frac {\partial e_{XC}(\rho(\vec{r}),\vec{\nabla} \rho(\vec{r}), \tau(\vec{r}))} {\partial \rho(\vec{r})}
                                              - \vec{\nabla} \frac {\partial e_{XC}(\rho(\vec{r}),\vec{\nabla} \rho(\vec{r}), \tau(\vec{r}))} {\partial \vec{\nabla} \rho(\vec{r})}\\
                                              +& \int \frac {\partial e_{XC}(\rho(\vec{r'}),\vec{\nabla} \rho(\vec{r'}), \tau(\vec{r'}))} {\partial \tau(\vec{r'})} 
                                             \frac {\delta \tau [\rho] (\vec{r'})}{\delta \rho(\vec{r})} d\vec{r'}
\end{aligned}
\end{equation}
where the third term is obtained with the functional derivative chain rules.
%\textcolor{red}{[What does the  prime mean here? Is the last term correct?]}
%\textcolor{blue}{[Eq. Edited to match the Eq. in Mark Gordon's paper]}
Typically, an exchange-correlation functional is implemented in quantum chemistry software in such a way that $\frac {\partial e_{XC}(\rho,\vec{\nabla} \rho, \tau)} {\partial \rho}$, $\frac {\partial e_{XC}(\rho,\vec{\nabla} \rho, \tau)} {\partial \vec{\nabla} \rho}$, and $ \frac {\partial e_{XC}(\rho,\vec{\nabla} \rho, \tau)} {\partial \tau}$ are returned from %functionals.
subroutines.
The $\frac{\delta \tau[\rho](\vec{r})}{\delta \rho}$ in Eq. (\ref{dExcdrho}) can be calculated as $\frac{\delta \tau}{\delta \psi}  \frac{\delta \psi}{\delta \rho}$; however, computing $\frac{\delta \psi[\rho](\vec{r})}{\delta \rho}$ is difficult. 
It was suggested by Zahariev \textit{et al.} \cite{doi:10.1063/1.4811270} and Yang \textit{et al.} \cite{PhysRevB.93.205205} that the Hamiltonian matrix elements of the pure meta-GGA exchange-correlation potential can be written as follows, using integrations-by-parts:
\begin{equation}
\begin{aligned}
 \int & \psi_i (\vec{r}) \frac{\delta E_{XC}[\tau[\rho]]}{\delta \rho(\vec{r})} \psi_j(\vec{r}) d\vec{r} \\
 \approx & \frac 1 2 \int \frac{\delta E_{XC}[\tau]}{\delta\tau(\vec{r})}\vec{\nabla} \psi_i(\vec{r})\cdot \vec{\nabla}\psi_j(\vec{r})d\vec{r}.
\end{aligned}
\end{equation}
This approach of computing the Hamiltonian matrix elements is used for the meta-GGA implementation in the FLOSIC code.

\vfill %This is for adjusting padding between paragraphs.

\subsection{FLOSIC}

%Perdew-Zunger SIC.
%In the first application of the PZ-SIC, SIC was applied to an LSD approximation.
%The self-interaction correction is calculated with orbital by orbital basis.
%SIE in DFA arises from the incomplete cancellation of electron self-Coulomb energy by the self-exchange energy. Perdew and Zunger proposed a SIC method (PZ-SIC) where the self-interaction of the occupied orbitals are subtracted from the total energy one by one
%In PZ-SIC formalism, the exchange-correlation energy of single occupied orbital cancels and eliminates self-Coulomb energy and self-exchange energy from each occupied orbital 
%
%\cite{PhysRevB.23.5048}.
%Hence, the SIC approximated term to the exchange-correlation energy is defined as follows,
FLOSIC uses the PZ-SIC total energy expression that removes the self-interaction of the occupied orbitals on an orbital by orbital basis:
\begin{equation}\label{eqnESIC}
 E_{}^{SIC}[\rho_\uparrow,\rho_\downarrow] =  E_{}[\rho_\uparrow,\rho_\downarrow]-\sum_\sigma \sum_i^{N_\sigma}\Big(U[\rho_{i\sigma}]+E_{XC}[\rho_{i\sigma},0]\Big)
\end{equation}
where $\sigma$ is the spin index, $i$ is the orbital index, and $N_{i\sigma}$ is the number of orbitals for spin $\sigma$. $\rho_\uparrow$ and $\rho_\downarrow$ denote spin up and spin down electron densities.  $\rho_{i\sigma} = |\phi_{i\sigma}|^2$, where the $\phi_{i\sigma}$ are the Fermi-L\"{o}wdin orbitals (FLO). The FO are constructed from a transformation on the KS orbitals using Eq. (\ref{eq:fod}).  These are normalized, but not mutually orthogonal. L\"{o}wdin orthogonalization yields the FLOs.

%It was shown by Pederson \textit{et al.} that the orbitals that variationally minimize the PZ-SIC total 
%energy of Eq. (\ref{eqnESIC}) %in SIC calculations 
%satisfy conditions known as localization equations given as
%\cite{doi:10.1063/1.446959, doi:10.1063/1.448266} 
%(M. R. Pederson, R. A. Heaton, and C. C. Lin, J. Chem. Phys.80, 1972 (1984) and [13] M. R. Pederson, R. A. Heaton, and C. C. Lin, J. Chem. Phys.82, 2688 (1985).)
%\begin{equation}
% \langle \phi_{j\sigma}|V_{j\sigma}^{SIC}-V_{i\sigma}^{SIC}|\phi_{i\sigma}\rangle = 0
%\end{equation}
%where $\phi$ is a localized orbital.%and find the localized orbitals that variationally minimize the energy 
%The SIC term is then added to the density functional approximated (DFA) equation.
The DFA-SIC single particle equations are 
\begin{equation}
 (H_\sigma^{DFA} + V_{i\sigma}^{SIC})\phi_{i\sigma} = \sum_j^{N_\sigma} \lambda_{ji\sigma} \phi_{j\sigma}.
\end{equation}
These are satisfied self-consistently for a given choice of the FODs, following the approach of Ref. \cite{PhysRevA.95.052505}.  We use an SCF convergence tolerance of $10^{-6}$ Ha.

\section{Results and discussion} %FLOSIC Atomic Total Energy Comparison by Functionals}\label{resultatoms}
\subsection{Atoms: total energies, ionization energies, and electron affinities}\label{atomandip}
%\textcolor{green}{ Raja:- Yoh, I have been revising text substantially and we have to move/reorder the tables/figures}

%We discuss the total energy of FLOSIC calculations. 
The focus of this work is to give a comprehensive assessment of the results of FLOSIC-SCAN calculations.  To do that we compare these to corresponding results for FLOSIC-LDA, FLOSIC-PBE, and for the corresponding uncorrected DFA's.  %We have chosen three different non-empirical exchange-correlation functionals that belong to the lowest three 
%rungs of Jacob's ladder \cite{doi:10.1063/1.1390175} --- namely, the LDA, PBE, and SCAN functionals. 
%
%{\bf The LDA calculations used the PW92
%exchange-correlation functional.  Since GGA provides a more compact electron density compared to LDA,
%all FLOSIC calculations  are started from self-consistent PBE electron densities without SIC. 
%As noted above the SCAN functional is extremely demanding on numerical mesh, the integrals involving 
%the SCAN functionals require ultra-fine mesh. We have therefore used the numerical mesh that is 
%about 3 times as large as the default mesh used in the code.  This part is being repetitive. }
%\textcolor{red}{(elaborate SCAN mesh discussion here; some discussion is in comp. section.)}
%
%of error that may arise due to numerical quadrature in FLOSIC-SCAN calculations we have chosen 
%optimized positions of the FLOSIC-LDA Fermi-Lowdin orbital descriptors.
%We perform two SCF cycles using identical FOD positions to ensure that local orbitals are stable.
%
%All FLOSIC results presented here are from fully self-consistent calculations. % for a given set of Fermi-Lowdin orbital descriptors.

\begin{figure}[!htb]
\centering
\includegraphics[width=\columnwidth]{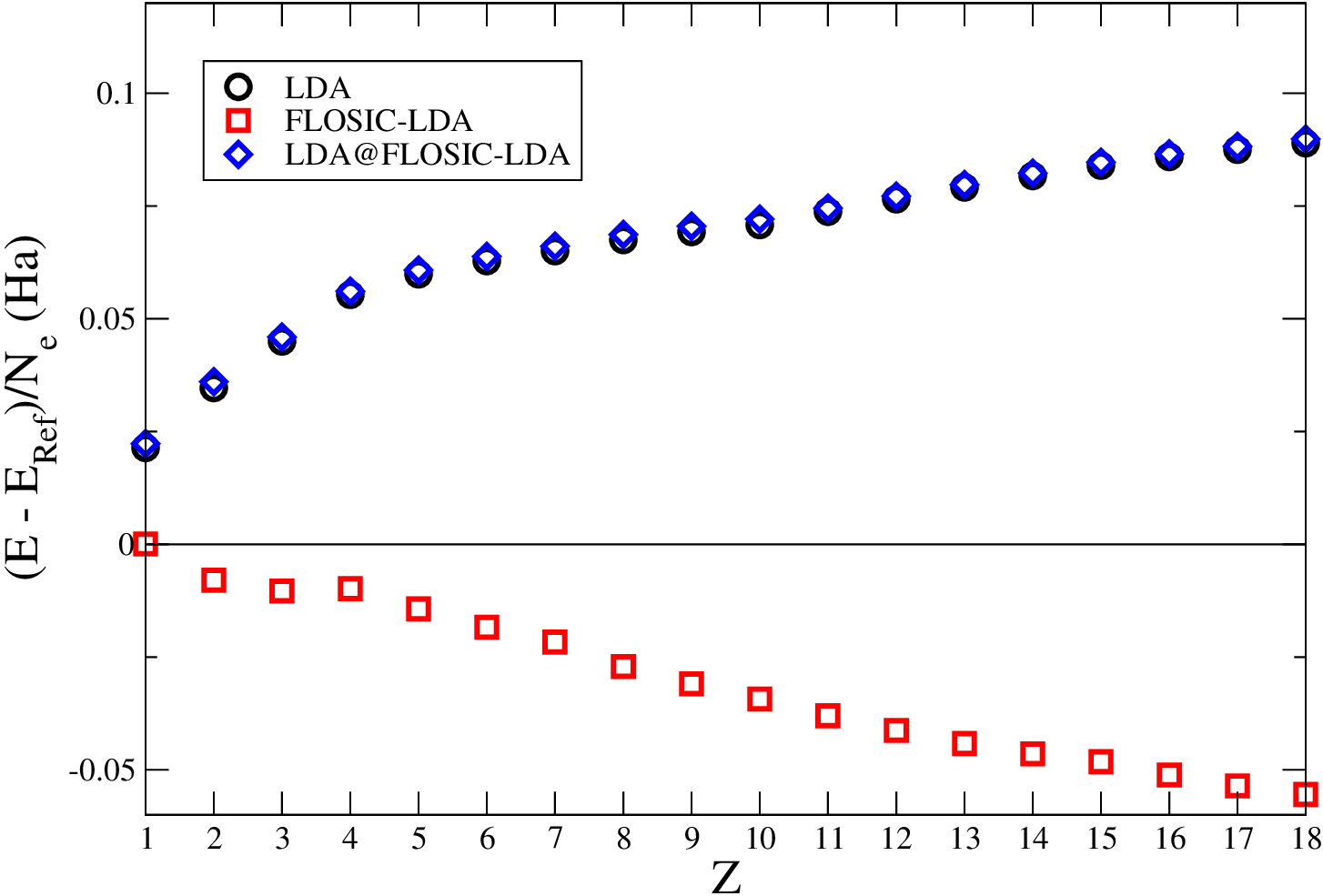}
\caption{Atomic total energies (in Ha) for LDA (black circles), FLOSIC-LDA (red squares), and LDA@FLOSIC-LDA (blue diamonds), compared against the reference values of Ref. \cite{PhysRevA.47.3649}. $(E - E_\text{Ref})/N_{e}$ is shown, where $N_e$ is the number of electrons. 
\label{ldas}}
%\vspace{20mm}
\end{figure}

\begin{figure}[!htb]
\centering
\includegraphics[width=\columnwidth]{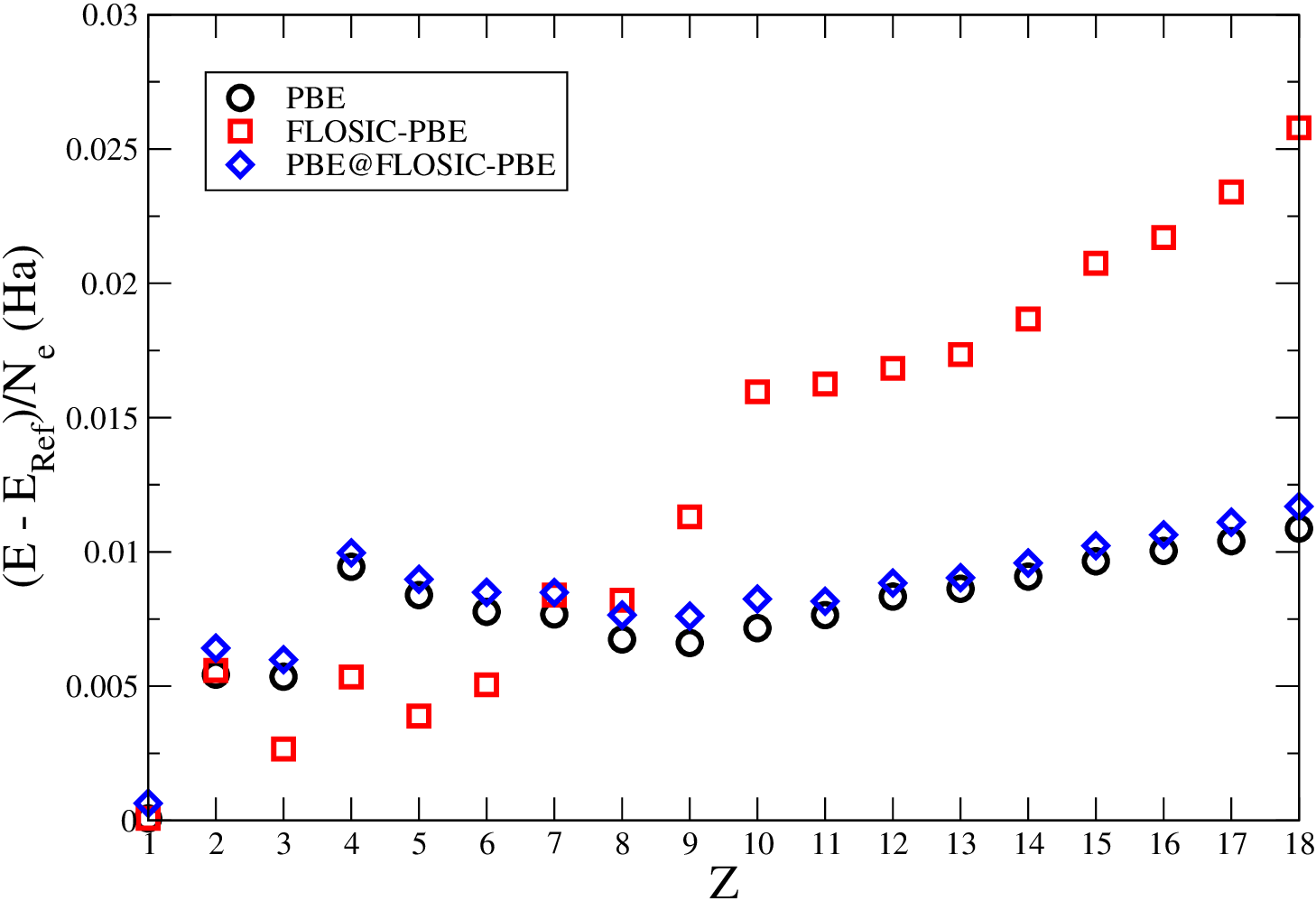}
\caption{Atomic total energies (in Ha) for PBE (black circles), FLOSIC-PBE (red squares), and PBE@FLOSIC-PBE (blue diamonds), compared against the reference values of Ref. \cite{PhysRevA.47.3649}. $(E - E_\text{Ref})/N_{e}$ is shown, where $N_e$ is the number of electrons. 
\label{pbes}}
%\vspace{20mm}
\end{figure}

\begin{figure}[!htb]
\centering
\includegraphics[width=\columnwidth]{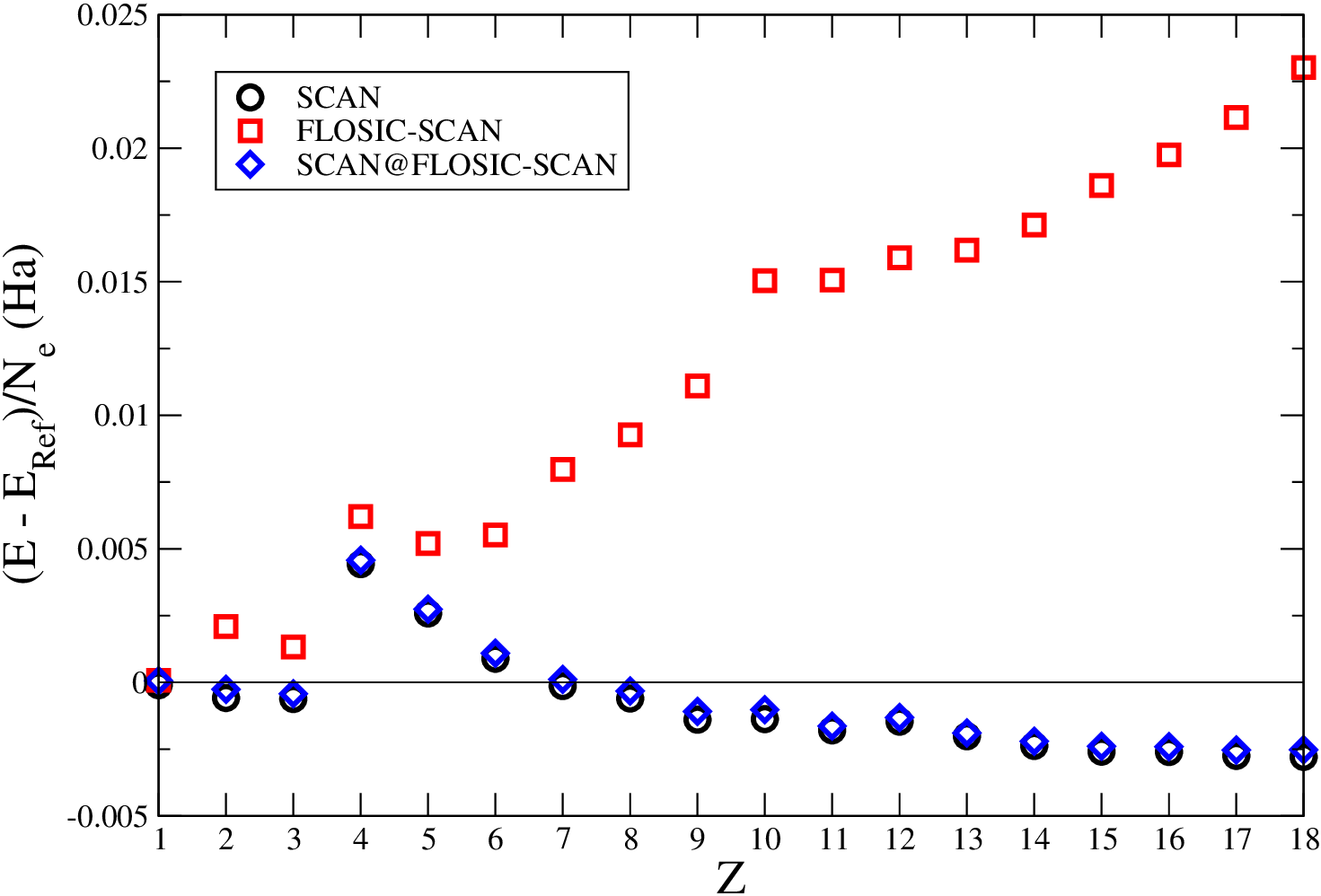}
\caption{Atomic total energies (in Ha) for SCAN (black circles), FLOSIC-SCAN (red squares), and SCAN@FLOSIC-SCAN (blue diamonds) compared against the reference values of Ref. \cite{PhysRevA.47.3649}. $(E - E_\text{Ref})/N_{e}$ is shown, where $N_e$ is the number of electrons.
\label{scans}}
%\vspace{20mm}
\end{figure}

The FLOSIC energies for atoms from H--Ar ($Z=1-18$) can be compared against accurate non-relativistic  total energies reported 
by  Chakravorty \textit{et al.} \cite{PhysRevA.47.3649}.
The deviation of the calculated total energies are given on a per electron basis as  $(E - E_\text{Ref})/N_e$, where $E$ is the 
FLOSIC energy, $E_\text{Ref}$ is the reference energy, %obtained through accurate calculations, 
and $N_e$ is the number of electrons in the given system.
The results are shown in Fig. \ref{ldas}--\ref{scans},
and the numerical errors of FLOSIC energies with respect to $E_\text{Ref}$ are presented 
in Table \ref{flosicstable}. 
%The mean absolute errors of the calculated total energies are 0.38 (FLOSIC-LSDA), 0.16 (FLOSIC-PBE), and 0.15 Ha (FLOSIC-SCAN).
As noted in earlier works \cite{doi:10.1063/1.4869581, PEDERSON2015153, doi:10.1063/1.4996498}, we find that the total energies with 
LSDA improve
within the FLOSIC method (shown in Fig. \ref{ldas}) %(Cf. Table S2)
  with a decrease in mean absolute error (MAE) from $0.73$ Ha (LSDA) to $0.38$ Ha (FLOSIC-LSDA).
  On the other hand, both PBE and SCAN total energies show a larger deviation when 
  corrected for self-interaction using FLOSIC as shown in Figures \ref{pbes} and \ref{scans}.
  The MAEs for total energy with PBE and FLOSIC-PBE are $0.083$ 
  and $0.159$ Ha respectively; for SCAN and FLOSIC-SCAN the MAE's are $0.019$ and $0.15$ Ha.  Thus, FLOSIC-PBE and FLOSIC-SCAN perform better than LSDA and FLOSIC-LSDA, but not as well as PBE and SCAN. 

   DFT calculation using accurate electron densities can eliminate density driven errors and give better energies \cite{VERMA201210,wasser-burke}.
   Since SIC restores the correct asymptotic behavior of the potential and one-electron self-interaction freedom
   %Since the FLOSIC method imposes additional exact constraints
   \cite{doi:10.1063/1.5090534}, it can provide a more physically reasonable density than a DFA calculation. 
   It is therefore of interest to calculate the total energies using the 
   self-consistent FLOSIC density in the standard GGA (PBE)
   and meta-GGA (SCAN) functionals. We denote these results as DFA@FLOSIC-DFA. For example,
   the SCAN@FLOSIC-SCAN is the result obtained by using 
   the self-consistent FLOSIC-SCAN electron density to evaluate the SCAN total energy.  
The DFA@FLOSIC-DFA with LDA, PBE, and SCAN produces atomic total energies that are very close to the self-consistent total energies of the respective DFA 
%with their parent functionals %DFA only total energies 
as shown in Figs. \ref{ldas}--\ref{scans}. 
%{\bf This points to the fact that 
%the correction to the density is very small with the application of the FLOSIC although the total 
%energy corrections are large in case of atoms. - but later we are saying the density is correct? } 
For completeness, we also tested the FLOSIC-LDA and FLOSIC-PBE densities in SCAN. The %error in 
SCAN@FLOSIC-SCAN, SCAN@FLOSIC-PBE and SCAN@FLOSIC-LDA energies 
 are very close, indicating that the respective FLOSIC densities are similar.  
%
%Similar results and
%conclusions can be drawn for the atomization energies and ionization potentials, discussed below.
%
%   
   Note that these DFA@FLOSIC-DFA results are obtained at no additional computational cost 
   beyond that of the FLOSIC calculations.
   %and that these total energies and atomization energies are comparable to
 %  those of the parent functionals.  %We emphasize %would to remind a reader %\textcolor{green}{'like to remind the reader'? } that the DFA@FLOSIC-DFA approach of obtaining good total energy estimate from the FLOSIC-DFA calculations is in a sense similar to schemes reported for scaling down the overcorrection in the FLOSIC-DFA/PZ-SIC calculations and was first used in Ref. \cite{DOEProceeding}. \textcolor{red}{[Check Ref.]}

\begin{figure}[!htb]
\centering
\includegraphics[width=\columnwidth]{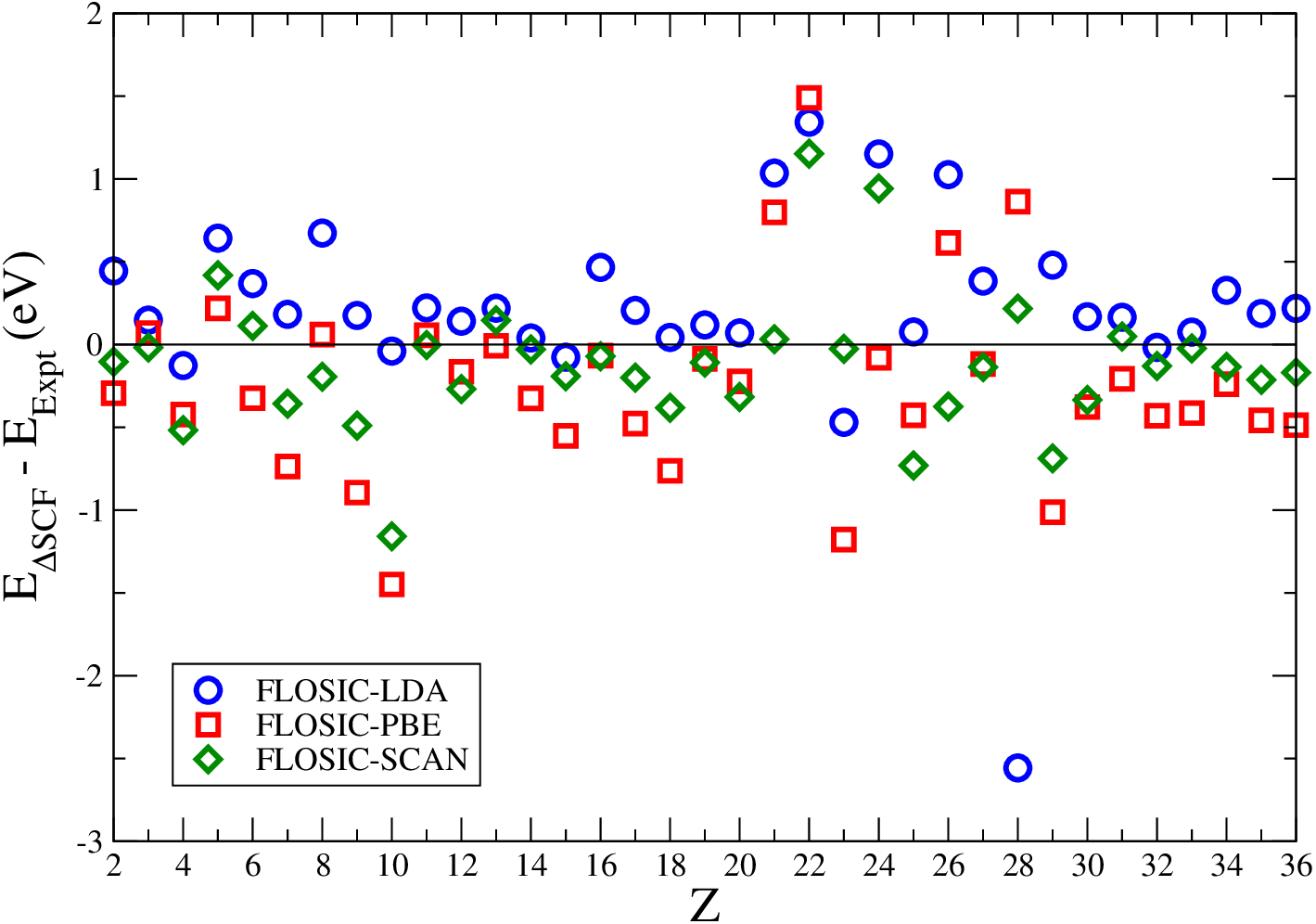}
\caption{Ionization energies (in eV) of atoms computed using FLOSIC-LDA (blue circles), FLOSIC-PBE (red squares), and FLOSIC-SCAN (green diamonds).
The energies are obtained by $\Delta$-SCF and compared against the experimental values of Ref. \cite{NIST_ASD}.}
\label{figureip}
%\vspace{20mm}
\end{figure}

   We also calculated the ionization potentials (IPs) for H--Kr atoms with FLOSIC applied to the LDA, PBE, and SCAN functionals.
The FOD optimization of cations is performed independently, and the resulting cation total energy $E_\text{cat}$ is then used to calculate the IP as 
\begin{equation}
E_\text{IP} =  E_\text{cat} - E_\text{neut}.
\end{equation}
%The FOD is optimized further if required.
The results from 
FLOSIC-LDA, FLOSIC-PBE, and FLOSIC-SCAN calculations
are summarized in Table \ref{errordeltascf}, and the energy differences from corresponding experimental energies \cite{NIST_ASD} are shown in Fig. \ref{figureip}. 
FLOSIC-LDA tends to overestimate the IPs with a few exceptions.
On the other hand, FLOSIC-PBE and FLOSIC-SCAN energies underestimate the experimental values.
%{\textcolor{red}{Note --- We should be looking at MAPE and not MAE for the overall performance.}}
%We obtain 
The mean absolute percentage errors (MAPE) in ionization energies are $7.68$, $5.13$, and $5.18$ \% %MAE of $0.619$, $0.397$, and $0.398$ eV 
for LDA, PBE, and SCAN, respectively. The MAPE values in IP are $5.01$, $5.04$, and $3.30$ \%   %MAE of $0.402$, $0.468$, and $0.448$ (LDA FOD) eV 
for FLOSIC-LDA, FLOSIC-PBE, and FLOSIC-SCAN respectively. For all three functionals, the values of IP are reduced overall with SIC compared to without. The IPs of PBE and SCAN are over-corrected with SIC.
This is seen in the sign of mean errors (ME); with SIC, the ME in IP changes from $0.342$ to $-0.230$ eV for PBE and from $0.277$ to $-0.278$ eV for SCAN.  In terms of mean absolute errors (MAE), FLOSIC improves the MAE for LSDA, from $0.619$ to $0.402$ eV, but increases it for PBE, from $0.397$ to $0.468$ eV.  The MAE is improved from $0.398$ to $0.299$ eV for SCAN and FLOSIC-SCAN. The results for LSDA and PBE are consistent with those of Vydrov and Scuseria \cite{doi:10.1063/1.1897378}. By comparing the SIC energy corrections for the neutral atoms and their cations, we observe that
the overcorrection of IPs with semilocal functionals occurs because the neutrals have a larger positive  correction than the cations, in most cases.
%Based on the MAPE values, the performance of LDA improves noticeably with FLOSIC calculations, while the performance of PBE and SCAN does not improve with FLOSIC due to the over-correction.
We point out that the optimization of the FOD at the level of the meta-GGA is important. We compared our FLOSIC-SCAN results with those calculated using descriptors optimized with FLOSIC-LDA. We find that the IPs with FLOSIC-SCAN
 %IP of FLOSIC-SCAN 
show a sizable improvement after performing FOD optimization. The MAE using FLOSIC-LDA optimized FODs is $0.448$ eV; this decreases to 
$0.299$ eV upon FLOSIC-SCAN optimization.  This reduction comes about in part by improving the Co IP.  Using FLOSIC-LDA FODs, the error for Co is $-5.082$ eV; using FLOSIC-SCAN FODs, the error drops to $-0.137$ eV.  %Similarly, the  MAPE  is high at $5.17$ \% with FLOSIC-LDA FODs,  %is reduced down from $5.17$ to $3.30$ \%.
%due to the large error of $-5.082$ eV for Co.  FOD optimization reduces this error to $-0.137$ eV, which contributes to improving the MAE to $3.30$ \%. 
This points to the importance of optimizing the FODs with a consistent functional. 

%In their PZ-SIC study of IP and electron affinities of atoms, Scuseria \textit{et al.} report that only for LSDA, $\Delta$-SCF IPs and EAs results are improved with PZ-SIC while the performance of other functionals --- BLYP, PBE, PBE0, and TPSS --- worsens with PZ-SIC \cite{doi:10.1063/1.1897378}.
%\textcolor{red}{Edit -  maybe we need DFT result?}
%\textcolor{blue}{[The trend is not same! Why are we quoting LDA-optimized FOD results since in the previous paragraph we are emphasizing the importance of optimization of FOD at the same level? If We consider the FLOSIC-SCAN IP, than MAE decreases - not same as Scuseria's results.]} 
% --> Text modified.
%We see a similar trend in our FLOSIC calculations where MAE of LDA improves from 0.619 to 0.402 eV with FLOSIC, and MAE of PBE increases from 0.397 to 0.468 eV.  
%For LDA and PBE, the FLOSIC results are consistent with PZ-SIC.
%0.398 to 0.448eV for SCAN with FLOSIC using LDA optimized FODs.
%In case of SCAN,
%as previously mentioned, we observed FOD optimization improves the IP of FLOSIC-SCAN where MAE decreases from 0.398 to 0.299 eV with SIC; however, the overcorrection can still be seen.
%This indicates that FOD optimizations are needed for depicting an accurate picture of an electron removal.

%\textcolor{red}{@calculation 1 - IP}
%
%We also performed DFA calculations using FLOSIC density refered as DFA@FLOSIC-DFA calculation. 
Similarly to what we have done for the total energy of atoms, we performed DFA@FLOSIC-DFA calculations for the IP.
%DFA calculations using a semilocal functional perform better in terms of ground state energies than SIC calculations with the same functional. 
%we expect these calculations to perform favorably when FLOSIC electron densities are incorporated.
%Since FLOSIC densities are more physically accurate, these calculations should provide an insights on whether the errors are from density driven error or functional driven error.
PBE@FLOSIC-PBE gives MAPE of $4.91$ \%, which is a smaller error than both PBE and FLOSIC-PBE. For SCAN@FLOSIC-SCAN (MAPE = $5.28$\%), we do not see a  performance improvement compared to SCAN ($5.18$\%) or FLOSIC-SCAN ($3.30$\%.) 

\begin{figure}[!htb]
\centering
\includegraphics[width=\columnwidth]{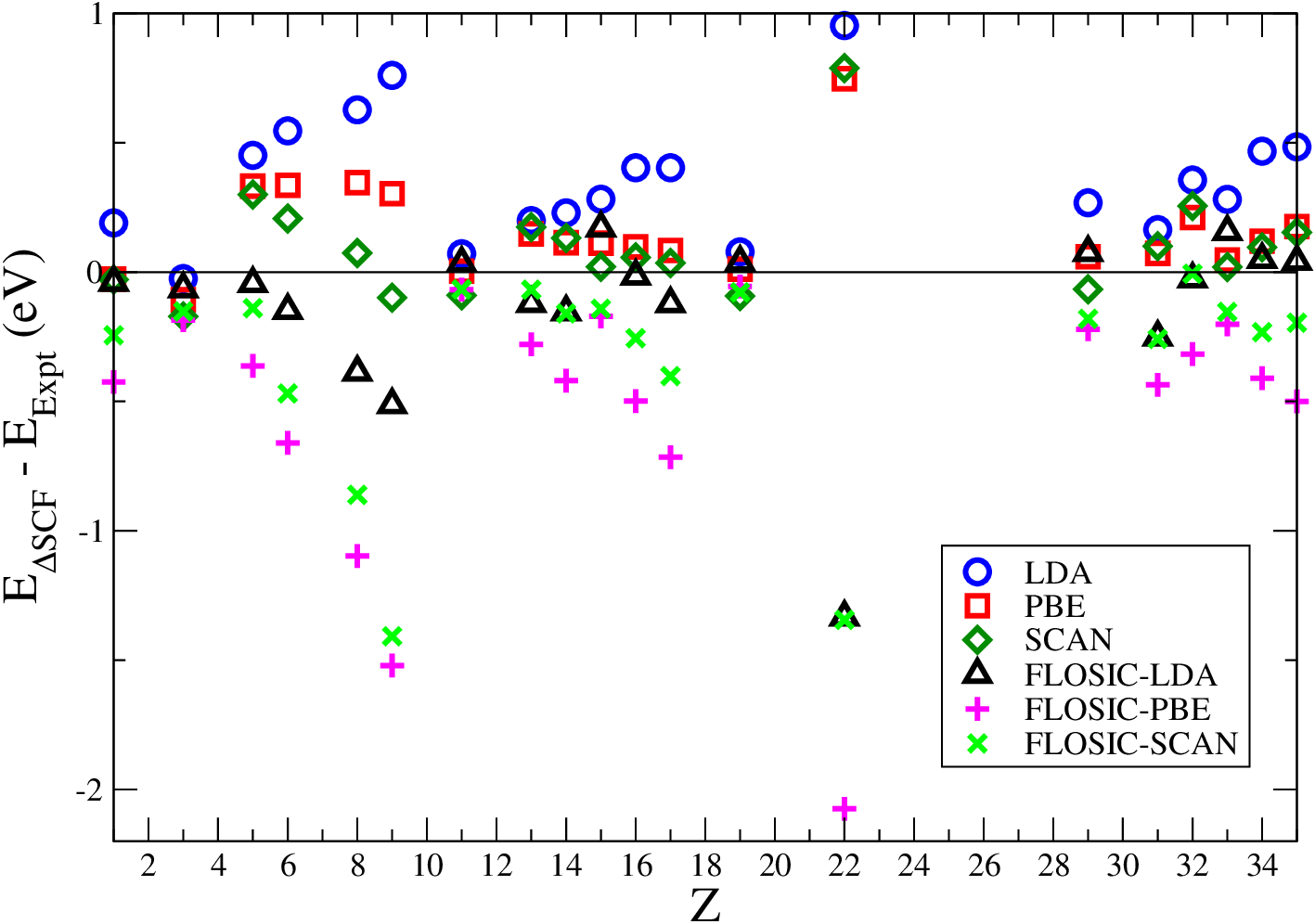}
\caption{\label{fig:flosicsa} Electron affinities (in eV) of 20 atoms computed using LDA (blue circles), PBE (red squares), SCAN(green diamonds), FLOSIC-LDA (black triangles), FLOSIC-PBE (magenta crosses), and FLOSIC-SCAN (green xs).  The energies are obtained by $\Delta$-SCF and compared against the experimental values of Ref. \cite{NIST_CCCBD}.}
\end{figure}

Finally, the electron affinities (EA) of the atoms were computed by taking the difference $EA = E_\text{neut} - E_\text{anion}$.
For the anion calculations, we added additional single Gaussian orbitals (s, p, and d-type) to the default NRLMOL basis set to account for the more diffuse nature of the anion wave functions. These extra orbitals share the same Gaussian exponents that are obtained using the relation $\beta(N+1) = \beta(N)^2/\beta(N-1)$ where $\beta(N)$ is the N-th Gaussian exponent in the basis. 
We computed EAs for H, Li, B, C, O, F, Na, Al, Si, P, S, Cl, K, Ti, Cu, Ga, Ge, As, Se, and Br, for which experimental EA values are available in Ref. \cite{NIST_CCCBD}.
In all the DFA anion calculations, the orbital eigenvalue of the highest occupied orbital becomes positive due to SIE \cite{PhysRevB.23.5048}, implying that the fully charged anions are not truly bound in DFA. Despite this, we adopt the common practice of computing EA values by taking total energy difference of an atom and its anion via $\Delta$-SCF. These are listed in Table \ref{tabea} and are comparable to 
those reported by Vydrov and Scuseria \cite{doi:10.1063/1.2176608}. The application of SIC results in negative HOMO orbital energies, due to the improved description of the exchange potential in the asymptotic region.  FLOSIC-PBE and FLOSIC-SCAN generally underestimate the EAs as seen from ME and MAE as well as in Fig. \ref{fig:flosicsa}. Overall, the performance of FLOSIC-LSDA is the
best among the three FLOSIC-DFAs.

DFA@FLOSIC-DFA calculations were also performed for EA similarly to the IP calculations.
For all three functionals, the errors with respect to experimental values are noticeably reduced compared to the pure DFA calculations (cf. Table III). This suggests that density-driven errors may be particularly important in describing the EA.  %In both PBE@FLOSIC-PBE and SCAN@FLOSIC-SCAN, the data points are distributed closer to the reference data points than their DFA or FLOSIC data points. %Since the FLOSIC densities do not have the positive HOMO problem, DFA@FLOSIC-DFA is qualitatively better in terms of both energy and density for EA.

\subsection{Atomization energies}\label{atomizationenergymol}
%We discuss FLOSIC performance on the selected molecules.
FLOSIC-LDA, -PBE, and -SCAN are also used to calculate  the total and atomization energies (AE) of a set of 
%molecules.
% are presented in this section. 
%The FLOSIC calculations are performed using LDA (PW92), PBE, and SCAN exchange-correlation functionals. 
%
%\subsection{Total Energy}
%The FLOSIC calculation is performed on a subset of 
$37$ molecules.  This supplements the FLOSIC-SCAN results that appeared recently \cite{doi:10.1063/1.5087065}.
Most of the molecules are taken from the G2/97 test set \cite{doi:10.1063/1.460205}; in addition, we include the six molecules from the AE6 test set \cite{doi:10.1021/jp035287b}, as well as HBr, LiBr, NaBr, FBr, Br$_2$, and cyclopentadienyl. Most of the geometries for these molecules were optimized using B3LYP with the 6-31G(2df,p) basis \cite{NIST_CCCBD}. 
 The geometries for  
O$_2$, CO, CO$_2$, C$_2$H$_2$, Li$_2$, CH$_4$, NH$_3$, and H$_2$O
 were optimized using the PBE functional and the default NRLMOL basis set.
FOD positions were initially optimized using FLOSIC-LDA and further optimized for FLOSIC-SCAN.
% and the identical FOD positions are used to compute FLOSIC-PBE and FLOSIC-SCAN self-consistently. {\bf  No further FOD optimization is performed in FLOSIC calculations with PBE and SCAN functionals.}
%The deviation of the total energies from experimental values  with LDA, PBE, and SCAN at the DFT only level and FLOSIC level are shown
%in figs \ref{X,Y,Z} respectively. 
% are shown in Table \ref{molenergy}.
%The majority of the systems are calculated with spin-polarized calculation.
%In some molecules, the system names are denoted with upo (pol) to indicate that the calculation is spin unpolarized (polarized).

%\subsection{Atomization Energy of molecules}
%We calculate the atomization energies of the selected systems.
The atomization energy of a molecule is defined as
\begin{equation}\label{eqnatomization}
E_\text{a}= \sum_i^{N_\text{atom}} E_i - E_\text{mol} > 0
\end{equation}
where $E_i$ is the energy of individual atoms, $N_\text{atom}$ is the number of atoms in the given molecule, and $E_\text{mol}$ is the total energy of a molecule.
Table \ref{errorAtomization} summarizes the errors in calculated AEs for DFA only, FLOSIC-DFA, and DFA@FLOSIC-DFA calculations. 
The experimental energies are taken from Ref. \cite{NIST_CCCBD}. %\cite{NISTatenergy}.
%and the results are shown in Figure \ref{figatomization}.
%Calculated errors are shown in Table \ref{errorAtomization}.
The MAEs are $99.0$, $65.7$, $196.0$, $84.3$, and $73.7$ kJ/mol
%The MAPE in the calculated atomization energies of those molecules are $8.64$, $5.22$, $13.42$, $9.67$, and $10.24$ \%
%, MAPE are $8.64$, $5.22$, $13.42$, $9.67$, and $10.45$ \%, and RMS are $146$, $102$, $321$, $114$, and $132$ kJ/mol 
for PBE, SCAN, FLOSIC-LDA, FLOSIC-PBE, and FLOSIC-SCAN respectively.
At the DFA level, SCAN performs much better than PBE resulting in the smallest MAE of 65.7 kJ/mol and MAPE of 5.22 \% among all five cases.
%We find that the FLOSIC calculations do not perform well overall.
On the other hand, FLOSIC-PBE and FLOSIC-SCAN results are generally worse than those of their parent functionals. 
%
%Among the five approaches, 
We find that FLOSIC-LDA performs the worst of the above five cases with overestimated  AE for many systems and  especially for Br$_2$ for which the MAPE is %showing the highest MAPE of 
13.42 \%.
%especially for Cyclobutane, Benzene, 2-Butyne, and Furan   %<== this is MAE
%%C$_4$H$_8$, Benzene, CH$_3$CCCH$_3$, and C$_4$H$_4$O 
%where the atomization energies are overestimated 
%by more than 600 kJ/mol.
FLOSIC-PBE and FLOSIC-SCAN atomization energies have similar MAEs and MAPEs.
%In both FLOSIC PBE and FLOSIC SCAN cases, the energy is lowered by SIC and underestimated overall.
%The result shows that DFT SCAN gives the smallest MAE among the calculations.
%e
It is interesting to note that for FLOSIC-SCAN, the %\textcolor{red}{initially LDA FODs are used, and the FODs are then optimized.} 
MAE is 94.5 kJ/mol using LDA-optimized FODs and it improves  to 73.7 kJ/mol after FOD optimization in FLOSIC-SCAN, indicating again that it is important to optimize FODs at a consistent level of theory.
%however, MAPE shows only small improvement from 10.5 to 10.2 \% implying that atomization energy may not be very sensitive to FOD placements.

%J\'{o}nsson's group reported that the atomization energy of PZ-SIC using PBE improves when complex orbitals are incorporated; however, 
%PZ-SIC using TPSS and B3LYP functionals perform worse on atomization energy calculation than pure DFA calculations even when complex orbitals are 
%used \cite{doi:10.1021/acs.jctc.6b00622}. Unlike total energy for atoms, atomization energy of molecules does not necessarily 
%always improve with the inclusion of complex orbitals. 
%It would be interesting to see how SCAN performs with complex orbitals.

\begin{figure}[!htb]
\centering
 \includegraphics[width=0.7\columnwidth]{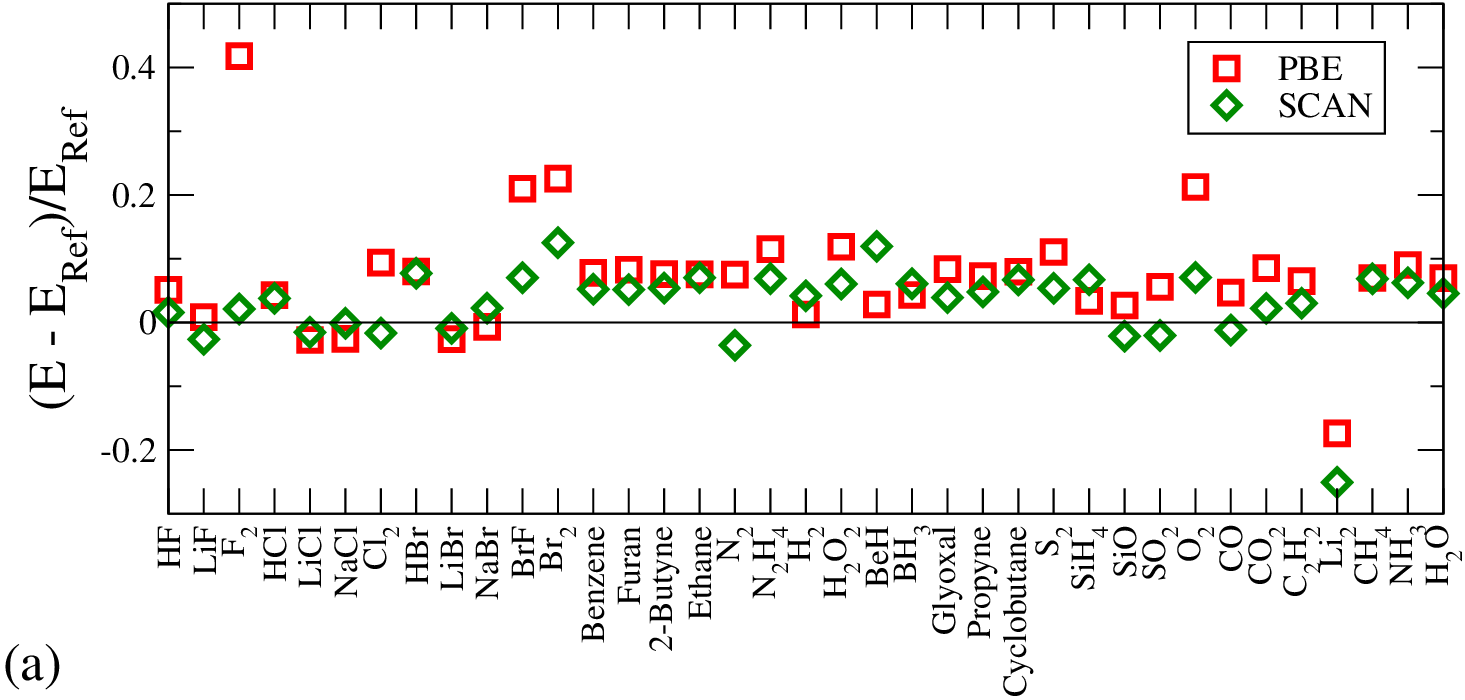}\\
  \vspace{10mm}
 \includegraphics[width=0.7\columnwidth]{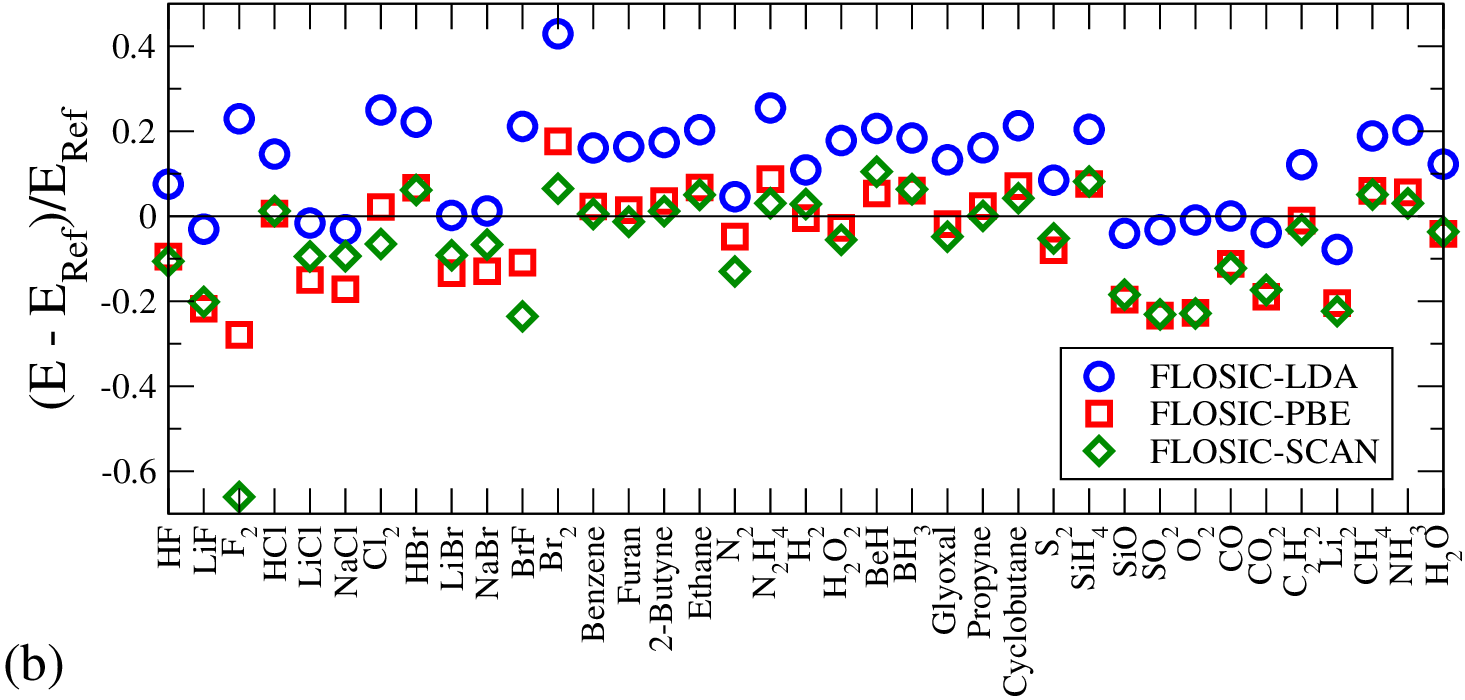}\\
  \vspace{10mm}
 \includegraphics[width=0.7\columnwidth]{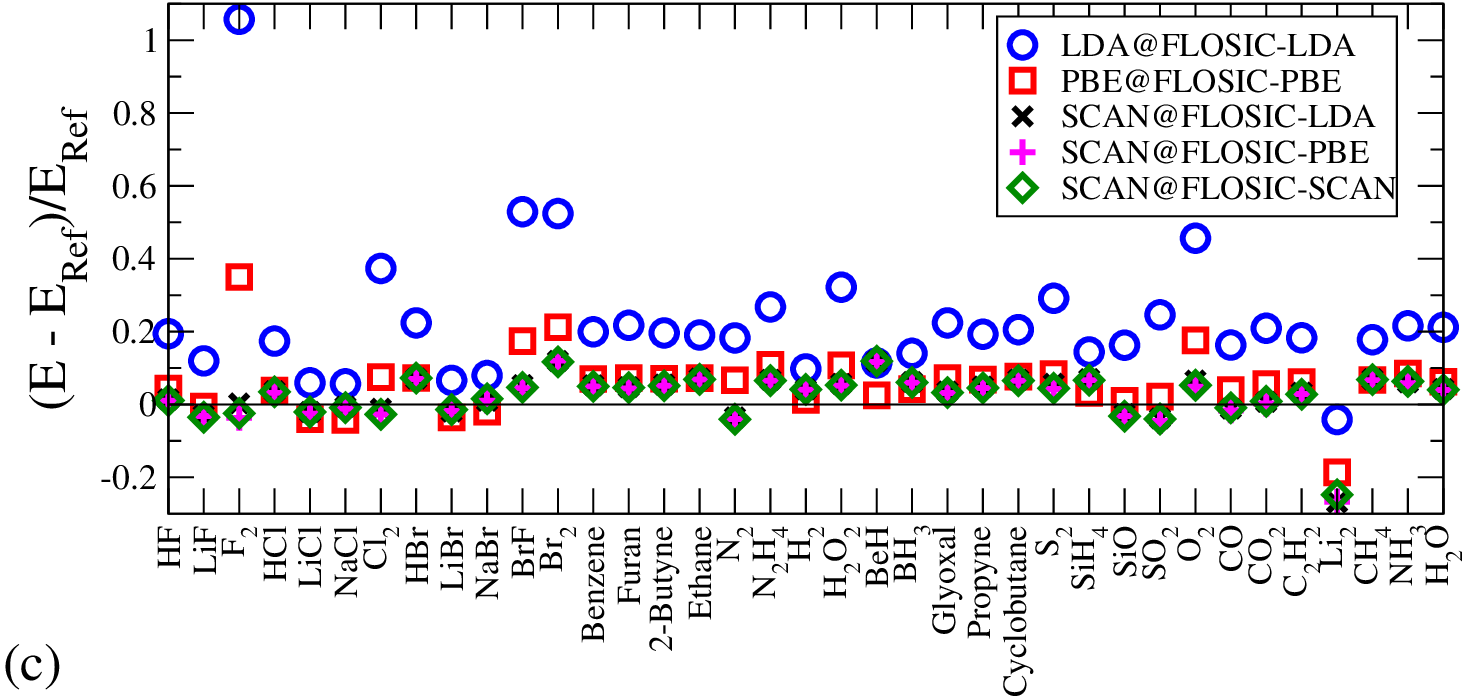}\\
\caption{Atomization energies of molecules compared against reference experimental values found in Ref. \cite{NIST_CCCBD}. $(E - E_\text{Ref})/E_\text{Ref}$ is shown: 
 (a) DFA, (b) FLOSIC, and (c) DFA@FLOSIC-DFA.
% (a) DFA calculations: PBE (red squares) and SCAN (green diamonds).
% (b) FLOSIC calculations: LDA (blue circles), PBE (red squares), and SCAN (green diamonds).
 %(c) DFA@FLOSIC-DFA calculations: LDA@FLOSIC-LDA (blue circles), PBE@FLOSIC-PBE (red squares), SCAN@FLOSIC-LDA (black Xs), SCAN@FLOSIC-PBE (magenta crosses), and SCAN@FLOSIC-SCAN (green diamons).
 \label{figatomization}}
 % \vspace{5mm}
\end{figure}

%In the literature, SIE is known to overemphasize electron correlation effects where the density is contracted towards the bonding regions \cite{doi:10.1080/00268970110111788}.
%DFA underestimates bond-length alternation due to SIE \cite{doi:10.1063/1.2047447}.
%It is known that the equilibrium bond lengths become shorter compared to the experimental bond lengths with the inclusion of SIC \cite{doi:10.1063/1.2176608, PhysRevA.55.1765}.
 Application of SIC generally results in an underestimation of the AEs compared to uncorrected DFA calculations (see Fig. \ref{figatomization}).  This is similar to results seen previously for semilocal functionals \cite{doi:10.1063/1.2176608}. In the FLOSIC calculations with semilocal functionals, we observe that SIC treatment raises the total energies of the molecules more than it raises the combined total energies of separated atoms with a few exceptions.
%SIC-PBE, SIC-TPSS, and SIC-PBEh are known to underestimate atomization energies significantly \cite{doi:10.1063/1.2176608}. Similar trend can be seen in our FLOSIC result from Fig. \ref{figatomization}.
%One of the reasons of this underbinding of molecules may come from the fact that bond lengths are not at equilibrium, which overestimates energies of molecules and reduces computed atomization energies.
 This observation was also noted by Shahi \textit{et al.} for real localized SIC orbitals \cite{doi:10.1063/1.5087065}. Consequently, the SIC treatment lowers atomization energies according to Eq. (\ref{eqnatomization}).
%This is one of the main reasons why the atomization energies are underestimated with FLOSIC with semilocal functionals.

%DFA@FLOSIC-DFA calculations
%\textcolor{red}{@calculation 2 - atomization energy}
%
%Since total energies of the molecules are overestimated using FLOSIC in case of semilocal functionals, parent DFA calculations have an advantage in performance of atomization energy.
  %As SIC improves the asymptotic description of potential and density, it is interesting  to examine how well DFA@FLOSIC-DFA performs for the AEs.  We 
We find that DFA@FLOSIC-DFA improves atomization energies with respect to \textit{both} the parent DFA and FLOSIC-DFA calculations.
The MAPE in AE for PBE is $8.64\%$ while that for FLOSIC-PBE is $9.67\%$. The MAPE for PBE@FLOSIC-PBE, on the other hand, is considerably smaller $7.72\%$. Similar improvement is also observed for SCAN. The MAPE of SCAN@FLOSIC-SCAN (5.05\%) is smaller than both the FLOSIC-SCAN ($10.24$ \%) and SCAN ($5.22$ \%). %These results show that FLOSIC can yield good AE if they are obtained using the FLOSIC density in the parent functional (DFA@FLOSIC-DFA). 
%For PBE, MAPE improves from $8.64$ to $7.72$ \% by using the improved density. 

%\subsection{Barrier heights of chemical reactions}
%The performance of SIC-SCAN in chemical reactions appeared in the publication before \cite{doi:10.1063/1.5087065}; however, there has not been a report of the FLOSIC-SCAN in chemical reactions.
%We use BH6 data set, which includes three chemical reactions: OH + CH$_4 \rightarrow$ CH$_3$ + H$_2$O, H + OH $\rightarrow$ H$_2$ + O, and H + H$_2$S $\rightarrow$ H$_2$ + HS \cite{doi:10.1021/ct600281g}. The geometries are as provided by the data set, and FODs are optimized.
%The total energies at left, right, and saddle points of those reactions are computed and the barrier heights of both forward and backward reactions are obtained. The results are compared against the reference values in Ref. \cite{doi:10.1021/jp035287b}, and they are shown in Table \ref{tableBH6}.

%In case of LDA, PBE, and FLOSIC-LDA, the barrier heights are all underestimated; this can be seen from ME and MAE. SCAN, on the other hand, underestimates two reactions and overestimate one reaction.
%For LDA, FLOSIC improves the performance where the MAE 
%Interestingly, both PBE and SCAN, MAE is reduced to almost half when FLOSIC is used --- MAE of PBE decreases from $8.0$ to $4.39$ kcal/mol, and that of SCAN decreases from $6.28$ to $3.17$ with FLOSIC. This may be because FLOSIC improves the transition states where the DFAs with semilocal functionals don't perform well due to stretched bonds.

\subsection{Dissociation energies}\label{sec:sie11}
We use SIE11 and SIE4$\times$4 test sets \cite{doi:10.1021/ct900489g,C7CP04913G}, sets of benchmark reactions that are known to be sensitive to self-interaction errors,
to investigate the performance of FLOSIC-SCAN on the dissociation energy calculations. The SIE11 test set consists of 11 systems that are directly affected from SIE.
The SIE4$\times$4 set consists of 4 positively charged dimers
(H$_2^+$, He$_2^+$, (NH$_3$)$_2^+$, and (H$_2$O)$_2^+$) separated at four different distances $R$ from the equilibrium distances $R_e$ ($R/R_e$=1.0, 1.25, 1.5, and 1.75); this set is designed to capture the effects of pure one-electron SIE.
Previously, Sharkas \textit{et al.} studied both SIE11 and SIE4$\times$4 with FLOSIC-LDA and FLOSIC-PBE and found that removal of self-interaction improves the performance in both case \cite{doi:10.1021/acs.jpca.8b09940}.
The dissociation energy is given as the difference of the complex total energy $E(X)$ and the fragments $E(X^+)$ and $E(X_2^+)$ as
\begin{equation}
    E_D = E(X) + E(X^+) - E(X_2^+).
\end{equation}
The results are compared against the reference values in Ref. \cite{doi:10.1021/ct900489g} and are shown in Table \ref{tableSIE}.
For LDA and PBE, we find MAE decreases from DFA to FLOSIC. The DFA calculations overestimate the total energies of both complexes and fragments, and it leads to large errors in the dissociation energies. FLOSIC is able to correct the total energies and improves errors in dissociation. This is expected since a removal of SIE should improve the results.
SCAN has relatively small self-interaction compared to other functionals, and DFA-SCAN shows smaller MAE in SIE11 ($10.4$ kcal/mol) than that for FLOSIC-LDA ($11.7$ kcal/mol).
In those data sets, the SIC treatment improves the performance of SCAN.
We find that FLOSIC-SCAN (MAE $= 5.7$ kcal/mol for SIE11 and $2.2$ kcal/mol for SIE4$\times$4) performs very well among the three functionals under both DFA and FLOSIC.

The SIE11 set is divided into five positively charged cationic and six neutral systems. 
DFA@FLOSIC-DFA calculations improve the errors for the neutral systems. This implies that those neutral systems are susceptible to density driven errors. 
For the SIE11 cationic systems, on the other hand, the MAEs of DFA@FLOSIC-DFA fall between FLOSIC and DFA indicating that full SIC treatment is needed. 
%This may be more to do with a functional driven error?
We observed the similar results for SIE4$\times$4 where full SIC is required as this dataset contains strechted bonds.

\subsection{Eigenvalues of the highest occupied orbitals}\label{evaluehoorbitals}

\begin{figure}[!htb]
\centering
\includegraphics[width=\columnwidth]{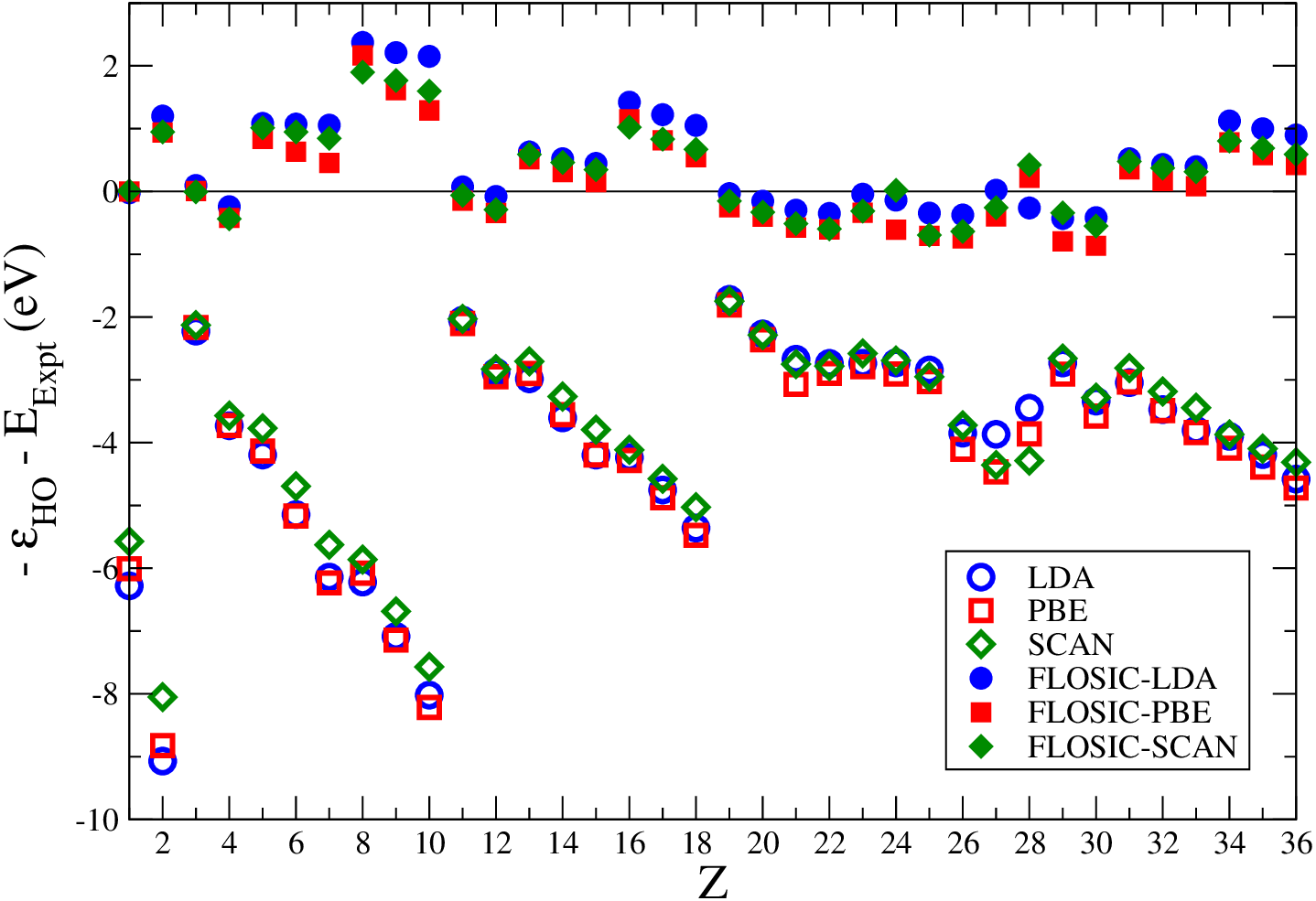}
\caption{Deviation of $-\varepsilon_{HO}$  from the corresponding experimental ionization potential\cite{NIST_ASD} (in eV) for atoms with $Z = 1 - 36$.  LDA (blue circles), PBE (red squares), SCAN (green diamonds), FLOSIC-LDA (filled blue circles), FLOSIC-PBE (filled red squares), and FLOSIC-SCAN (filled green diamonds) values are shown.}
\label{figureeho}
%\vspace{20mm}
\end{figure}

   In exact DFT, the negative of 
   the highest occupied eigenvalue equals the first ionization energy of the system
   \cite{PhysRevA.30.2745, PhysRevB.60.4545}.
   %\textcolor{red}{cite Perdew-Sahni \textit{et al.} PRL84, and Harbola 94 or 95}. 
   This property has been widely used to adjust the magnitude of the exchange 
   potential or exact exchange potentials in practical DFT calculations \cite{PhysRevLett.105.266802, doi:10.1021/ct5000617}.
%The highest occupied orbital eigenvalues $\varepsilon_{HO}$ of atoms AN$=1-36$ are also compared against experimental electron removal energies (Fig. \ref{figureeho}). 
In Fig. \ref{figureeho} we compare the SCAN and FLOSIC-SCAN HOMO orbital eigenvalues $\varepsilon_{HO}$ of atoms $Z=1-36$ against experimental electron removal energies. We also include the corresponding results for LDA, PBE, FLOSIC-LDA, and FLOSIC-PBE for comparison.
Table \ref{errorEhomo} shows that the MAEs of DFA orbital eigenvalues are $4.06$, $4.15$, and $3.88$ eV 
for LDA, PBE, and SCAN respectively, and MAE of FLOSIC-DFA eigenvalues are $0.67$, $0.59$, and $0.61$ eV 
%ME of FLOSIC eigenvalues are $0.49$, $0.19$, and $0.32$ eV
in the same order. 
Although the size of the errors of the HOMO eigenvalues is similar to the errors in IP calculated using total energy differences, the corrections to 
the HOMO
eigenvalues are much larger.  
The DFA HOMO eigenvalues significantly underestimate the electron removal energies for all three functionals. FLOSIC corrects 
this and reduces the MAE by a factor of 6 to 7.

\begin{figure}[!htb]
\centering
\includegraphics[width=\columnwidth]{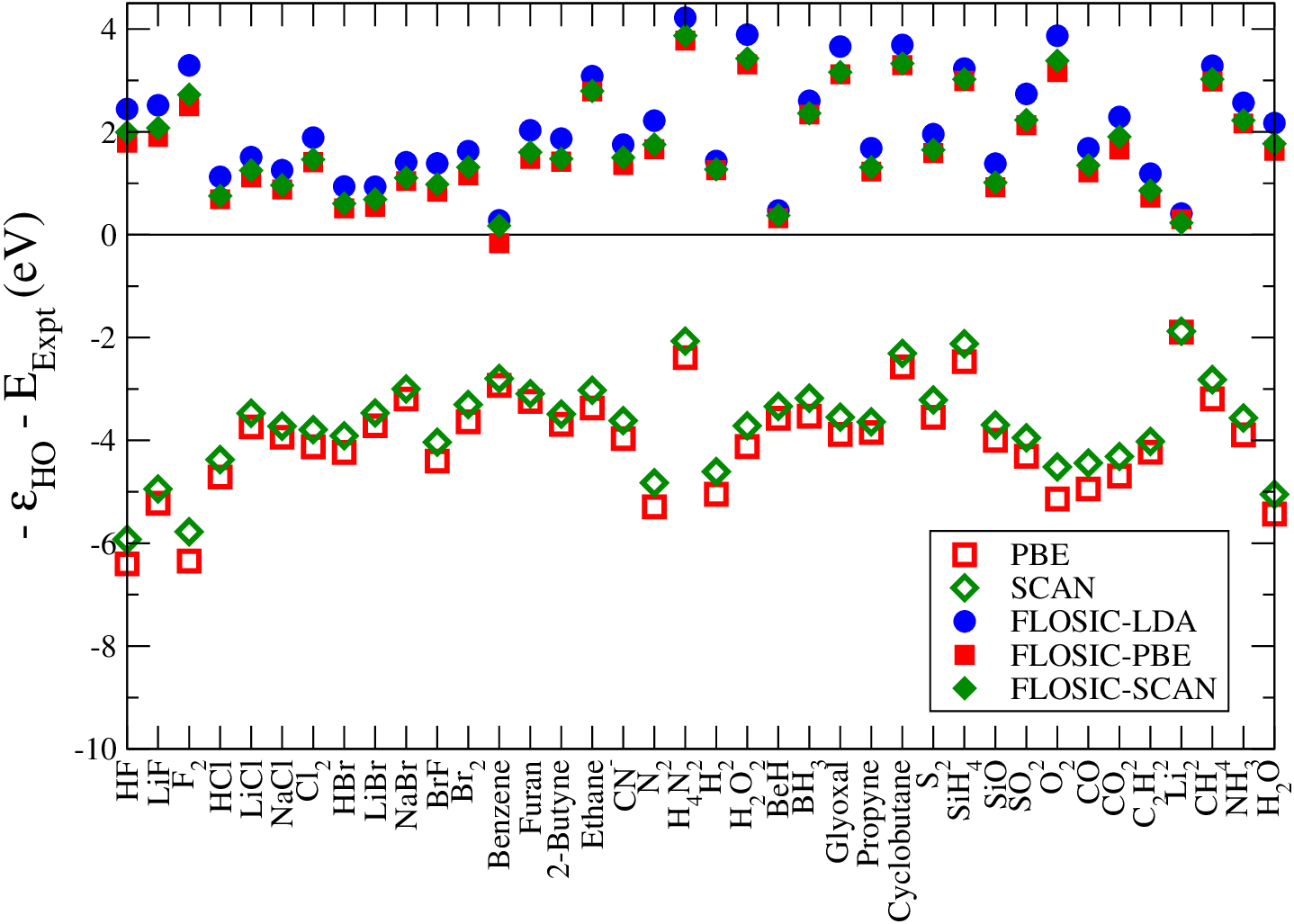}
\caption{Deviation of $-\varepsilon_{HO}$ from the corresponding experimental ionization potential (in eV) for a test set of molecules. The experimental values are from  Ref. \cite{NIST_webbook} and Ref. \cite{HuberHerzberg1979}.
PBE (red squares), SCAN (green diamonds), FLOSIC-LDA (filled blue circles), FLOSIC-PBE (filled red squares), and FLOSIC-SCAN (filled green diamonds) values are shown.}
\label{figureehomol}
%\vspace{20mm}
\end{figure}

Similar improvement in the eigenvalues of the HOMO is also seen for the set of molecules studied here (Fig. \ref{figureehomol}). As with the atoms, the HOMOs for the molecules are too high, understimating electron removal energies.  In all cases, the HOMO eigenvalues are significantly lowered resulting in overestimated ionization potentials with FLOSIC. 
%\textcolor{red}{[Comment: does anyone know why the errors for $\varepsilon_{HO}$ is larger for molecules?]}
%\textcolor{blue}{[YY. Transition metals tend to have smaller errors and molecules studied here don't have any transition metals. Also closed systems seem to have a large error.]}

%   While the use of FLOSIC deteriorates the agreement between the PBE and SCAN  total energies 
%   with the highly accurate reference values, 
Eliminating self-interaction error 
   improves the description of the potential seen by the electrons in the asymptotic region. This accounts for the significant improvement 
   in the eigenvalue of the highest occupied orbitals as can be seen from Tables \ref{errorEhomo} and \ref{errorEhomoMol}.
  As HOMO eigenvalue is related to the asymptotic decay of the electron density 
   \cite{PhysRevA.32.2010, PhysRevA.30.2745},
  % \textcolor{red} {cite Levy and co-workers [Phys. Rev. A 32, 2010 (1985); 30, 2745 (1984)]}
  it is reasonable to expect that the FLOSIC electron density is more accurate in the valence region than the corresponding uncorrected DFA density. 

\section{Conclusion}
We implemented meta-GGA functionals in the FLOSIC code and compared the performance of FLOSIC-SCAN to that of FLOSIC-LDA and FLOSIC-PBE calculations for a variety of properties.   Total energies of atoms from H--Kr are obtained.
We find that SCAN performs well in the total energy calculations, however, correcting for self-interaction errors using 
FLOSIC worsens the total energies. As also has been noted in a few earlier PZ-SIC works,
the application of the FLOSIC method deteriorates the total energies and atomization energies where self-interaction
errors are small.  Only in the case of LDA, the removal of self-interaction errors improves the results 
over the parent DFA functional.  For ionization potentials, FLOSIC improves ionization potentials 
for LDA but worsens them for PBE and SCAN. 
A pragmatic solution to obtain meaningful estimates of the atomization and total energies is to 
compute these quantities using 
the self-consistent self-interaction corrected electron density and Kohn-Sham orbitals in the parent functional.
This peturbative procedure %unlike previously adopted approaches of scaling down SIC correction for obtaining good estimates of total and atomization energies, preserves the effect of the correct $-1/r$ asymptotic
%behavior of the potential on the density and  
does not require any additional computational effort beyond the FLOSIC calculation.
%Performance assessment of such a procedure has been carried out. 
Our  results show that the 
%in better 
%of 
total energies, atomization energies, electron affinities and ionization energies (using $\Delta$-SCF) obtained using 
such a procedure are of comparable quality as of their parent functionals while keeping the benefits from SIC such as physically accurate electron densities and improved occupied orbital eigenvalues.
For the SCAN functional, we saw some improvement over DFA-SCAN in total and atomization energies as judged from 
MAEs of these quantities. 
The procedure adopted here 
is similar to that used in removing delocalization errors (density driven errors) in 
the literature \cite{VERMA201210} and is expected to be more accurate for ionization potentials and electron 
affinities for larger systems. The present work shows that FLOSIC calculations can provide  accurate estimates of the near equilibirum properties (e.g. total and atomic energies) 
where SIE are small by employing  DFA@FLOSIC-DFA approach while providing 
accurate description of properties like dissociation energies (using full FLOSIC-DFA) 
where SIC errors are large. Alternative approaches to rectify the overcorrection of the PZ-SIC/FLOSIC methods are being pursued in our laboratory.

\newpage
\section*{Figures}

\newpage
%\begin{figure}[!htb]
%\centering
%\includegraphics[width=\columnwidth]{Figs/flosic.eps}
%\caption{Comparison of FLOSIC-LDA (black circles), FLOSIC-PBE (red squares), and FLOSIC-SCAN (blue diamonds) total energies with reference values \cite{PhysRevA.47.3649} (in Ha) for atoms with $Z=1-18$. $(E - E_\text{Ref})/N_{e}$ is shown, where $N_e$ is the number of electrons. 
%\label{flosics}}
%%\vspace{20mm}
%\end{figure}

\newpage

\newpage

\newpage

\newpage

\newpage

\newpage

\newpage

\newpage

%\begin{figure}[ht]
%\centering
%\includegraphics[width=\columnwidth]{Figs/dipolemoment.eps}
%\caption{\label{fig:diplole}Deviation of the dipole moment from experimental values for selected molecules (in Debye). Legend: PBE (red squares), SCAN(green diamonds), FLOSIC-LDA (blue circles), FLOSIC-PBE (magenta triangles), and FLOSIC-SCAN ( green xs).}
%\end{figure}

\newpage
\section*{Tables}
%main  Atom MAE
\begin{table}[!htb]
\begin{center}
\caption{Mean absolute error (MAE in Ha) of the total energies of atoms with $Z=1-18$ calculated with various methods when compared against reference values given in Ref. \cite{PhysRevA.47.3649}.}
\label{flosicstable}
\begin{ruledtabular}
\begin{tabular}{|c|c|} %updated 9/21/18 updated 2: 2/12/19
Method				&	MAE (Ha)	\\\hline
LDA				&	0.726125		\\
FLOSIC-LDA		&	0.380502		\\
LDA@FLOSIC-LDA	&	0.734249		\\\hline
PBE				&	0.082958		\\
FLOSIC-PBE		&	0.159131		\\
PBE@FLOSIC-PBE	&	0.089404		\\\hline
SCAN			&	0.019197		\\
FLOSIC-SCAN		&	0.147113		\\
SCAN@FLOSIC-SCAN	&	0.017547	\\
\end{tabular}
\end{ruledtabular}
\end{center}
\end{table}

%main   Delta SCF Summary
%Table for ME MAE of IP
\begin{table}[!htb]
\begin{center}
\caption{Deviation of calculated ($\Delta$-SCF) ionization potentials from experimental values for atoms $Z=2-36$ for several methods. Mean errors (ME, in eV), mean absolute errors (MAE, in eV), and mean absolute percentage errors (MAPE) are shown.
 %\textcolor{green}{Table captions of Tables II and III are a little inconsistent with IV and V. i.e. "Deviations of -- from -- . ME, MAE,... are presented/shown. All values are in eV/kJ/mol."}
} 
%\label{iptable}
\label{errordeltascf}
\begin{ruledtabular}
\begin{tabular}{|c|c|c|c|}
Method		&	ME 		&	MAE  	&	MAPE	\\
                &   (eV)    &   (eV)    &   (\%)    \\\hline
LDA				&	0.586	&	0.619	&	7.68		\\
FLOSIC-LDA		&	0.214	&	0.402	&	5.01		\\
LDA@FLOSIC-LDA			&	0.482	&	0.521	&	6.45		\\\hline
PBE				&	0.342	&	0.397	&	5.13		\\
FLOSIC-PBE		&	-0.230	&	0.468	&	5.04		\\
PBE@FLOSIC-PBE			&	0.272	&	0.372	&	4.91		\\\hline
SCAN			&	0.277	&	0.398	&	5.18		\\
FLOSIC-SCAN (LDA FOD)	&	-0.278	&	0.448	&	5.17	\\
FLOSIC-SCAN (Optimized FOD)&	-0.123	&	0.299	&	3.30	\\
%SCAN@FLOSIC-LDA		&	0.204	&	0.376	&	4.93	\\
%SCAN@FLOSIC-PBE		&	0.164	&	0.334	&	4.34	\\
SCAN@FLOSIC-SCAN (LDA FOD)	&	0.244	&	0.402 &	5.28	\\%0.40237	\\
SCAN@FLOSIC-SCAN (Optimized FOD)&	0.241	&	0.402	&	5.28 \\%0.40206 \\
\end{tabular}
\end{ruledtabular}
\end{center}
\end{table}

\newpage

\begin{table}[!htb]
\begin{center}
\caption{\label{tabea}%
Electron affinities of 20 atoms calculated with various methods and compared to experimental values \cite{NIST_CCCBD}. Mean error (ME) and mean absolute error (MAE) are shown, both in eV.}
\begin{ruledtabular}
\begin{tabular}{|l|c|c|}
\textrm{Method}&
\textrm{ME }&
\textrm{MAE}\\
%\textrm{MAPE}\\
% & (eV) & (eV) & (\%) \\
\colrule
LDA             & 0.359	    &   0.362    \\%& 85.66	\\
FLOSIC-LDA      & -0.133    &   0.189    \\%& 89.36  \\
LDA@FLOSIC-LDA  &  0.227    &   0.231    \\ \hline
PBE             & 0.159     &   0.172    \\%& 59.01  \\
FLOSIC-PBE      & -0.531    &   0.531    \\%& 158.13  \\
PBE@FLOSIC-PBE  &  0.038    &   0.080    \\ \hline
SCAN            & 0.093     &   0.148    \\%& 61.01  \\
FLOSIC-SCAN     & -0.341    &   0.341   \\%&   \\
SCAN@FLOSIC-SCAN &   0.031       &    0.126   \\
\end{tabular}
\end{ruledtabular}
\end{center}
\end{table}
%Note to myself - MAPE is excluded here since the percentage error of Ti skews everything (-2500% to 1000%). 

%main   Summary Atomization Energy
%Table for ME MAE RMS
\begin{table}[!htb]
\begin{center}
\caption{Atomization energies for the test set of molecules featured in Fig. \ref{figatomization}. Mean absolute errors (MAE, in kJ/mol), mean percentage errors (MPE), mean absolute percentage errors (MAPE), and root mean square errors (RMS, in kJ/mol) are shown.
}
%LDA@FLOSIC-LDA denotes the LDA results obtained using the FLOSIC-LDA density.} % in kJ/mol.}
\label{errorAtomization}
\begin{ruledtabular}
\begin{tabular}{|c|c|c|c|c|}
Method							&	MAE 			&	MPE		&	MAPE  	&	RMS	\\
							&      (kJ/mol)		&	(\%)		&	(\%)		&	(kJ/mol)\\\hline
FLOSIC-LDA					&	 195.95		&	11.93	&	13.42	&	321.16	\\
LDA@FLOSIC-LDA				&	 267.41		&	22.78	&	23.00	&	381.49	\\\hline
PBE							&	 98.99		&	7.24		&	8.64		&	146.48	\\
FLOSIC-PBE					&	 84.30		&	-4.81	&	9.67		&	114.21	\\
PBE@FLOSIC-PBE				&	 88.85		&	5.93		&	7.72		&	133.27	\\\hline
SCAN						&	 65.69		&	3.01		&	5.22		&	102.42	\\
FLOSIC-SCAN (LDA FOD)			&	 94.50		&	-4.84	&	10.45	&	131.78	\\
FLOSIC-SCAN (Optimized FOD)	&	 73.72		&	-6.78	&	10.24	&	97.83	\\
%SCAN@FLOSIC-LDA				&	 65.11 		&	2.61		&	5.14		&	102.25	\\
%SCAN@FLOSIC-PBE				&	 62.26		&	2.11		&	5.09		&	97.08	\\
SCAN@FLOSIC-SCAN (LDA FOD)	&	 63.38		&	2.31		&	5.10		&	98.82	\\
SCAN@FLOSIC-SCAN (Optimized FOD)&	 62.84		&	2.35		&	5.05		&	97.87	\\
\end{tabular}
\end{ruledtabular}
\end{center}
\end{table}

%%%%
\newpage

\begin{table}[!htb]
\begin{center}
\caption{\label{tableSIE}SIE11 and SIE4$\times$4 dissociation energies calculated by various methods and compared to reference values from Ref. \cite{doi:10.1021/ct900489g}. Mean absolute errors (MAE, in kcal/mol) of SIE11 (5 cationic, 6 neutral, and 11 combined systems) and SIE4$\times$4 are shown.}
\begin{ruledtabular}
\begin{tabular}{|l|c|c|c|c|}
%Method  &	MAE   & MAE  & MAE & MAE \\\hline
Method  &   SIE11, 5 cationic & SIE11, 6 neutral & SIE11  & SIE4$\times$4 \\\hline
LDA             &  22.9  &  13.4  &   17.8  & 27.5 \\
FLOSIC-LDA	    &  14.8  &   9.0  &   11.7  & 3.0  \\
LDA@FLOSIC-LDA  &  20.1  &   8.9  &   14.1  & 21.2 \\ \hline
PBE	            &  12.7  &  10.9  &   12.1  & 23.3 \\
FLOSIC-PBE	    &  8.9   &   6.4  &   7.5   & 3.4 \\
PBE@FLOSIC-PBE  &  9.6   &   4.5  &   7.2   & 15.1 \\ \hline
SCAN	        &  10.4  &   9.9  &   10.4  & 17.9 \\
FLOSIC-SCAN	    &  5.1   &   6.2  &   5.7   & 2.2 \\
SCAN@FLOSIC-SCAN & 8.8   &   4.9  &   6.9   & 12.4 \\
% sig. fig. is 1st decimal place
\end{tabular}
\end{ruledtabular}
\end{center}
\end{table}

%main   ehomo atom summary
%Table for ME MAE of IP
\begin{table}[!htb]
\begin{center}
\caption{Deviation of $-\varepsilon_{HO}$ from the corresponding experimental ionization potential for atoms with $Z = 1-36$. Mean errors (ME) and mean absolute errors (MAE) are given in eV.} 
\label{errorEhomo}
\begin{ruledtabular}
\begin{tabular}{|c|c|c|}
Method	&	ME 		&	MAE  		\\\hline
LDA		&	-4.059	&	4.059			\\
PBE		&	-4.150	&	4.150			\\
SCAN		&	-3.880	&	3.880			\\\hline
FLOSIC-LDA	&	 0.494	&	0.672			\\
FLOSIC-PBE	&	 0.189	&	0.590			\\
FLOSIC-SCAN (LDA FODs)	&	 0.314	&	0.622			\\
FLOSIC-SCAN (FOD optimized)&	0.318	&	0.606		\\
\end{tabular}
\end{ruledtabular}
\end{center}
\end{table}

%main   ehomo molecule summary
%Table for ME MAE of EHomo of Molecules
\begin{table}[!htb]
\begin{center}
\caption{Deviation of $-\varepsilon_{HO}$ from the corresponding experimental ionization potential for the set of molecules featured in Fig. \ref{figureehomol}. Mean errors (ME) and mean absolute errors (MAE) are given in eV.} 
\label{errorEhomoMol}
\begin{ruledtabular}
\begin{tabular}{|c|c|c|}
Method	&	ME 		&	MAE  		\\\hline
PBE			&	-4.023	&	4.023			\\
SCAN		&	-3.699	&	3.699			\\\hline
FLOSIC-LDA	&	 2.104	&	2.104			\\
FLOSIC-PBE	&	 1.658	&	1.667			\\
FLOSIC-SCAN (LDA FODs)	&	1.790 	&	1.790		\\
FLOSIC-SCAN (Optimized FOD)&	1.762	&	1.762		\\
\end{tabular}
\end{ruledtabular}
\end{center}
\end{table}

%\begin{table}
%\begin{center}
%\caption{\label{tabdipole}%
%Calculated dipole moments of the 34 molecules featured in Fig. \ref{fig:diplole} compared to experimental values.  The mean absolute errors for various methods are given in Debye.}
%\begin{ruledtabular}
%\begin{tabular}{lccc}
%\textrm{Method}&
%\textrm{MAE}\\
%\colrule
%PBE         &  0.080        \\
%SCAN        &  0.033        \\
%FLOSIC-LDA  &  0.101        \\
%FLOSIC-PBE  &  0.071        \\
%FLOSIC-SCAN &  0.095        \\
%\end{tabular}
%\end{ruledtabular}
%\end{center}
%\end{table}

\newpage
\vfill
\section*{Supplementary material}
See supplementary material for detailed results of the total energies, IP, EA, Atomization energies, and $-\varepsilon_{HO}$ for the systems studied in this manuscript and detailed results for SIE11, SIE4$\times$4, and BH6 molecular test sets.

\begin{acknowledgments}
%The authors are supported by DOE (\textcolor{red}{insert project number})
The authors gratefully acknowledge discussions with Profs. Mark R. Pederson, John Perdew, Jianwei Sun, and  Dr. Jorge Vargas.
	The initial phase of the work (implementation of meta-GGAs) was supported by the Office of Basic Energy Sciences, U.S. Department of Energy DE-SC0002168 and DE-SC0006818 while the applications using FLOSIC are supported by DE-SC0018331 as a part of the Computational Chemical Sciences program.
	This research used resources of the National Energy Research Scientific Computing Center (NERSC), a U.S. Department of Energy Office of Science User Facility operated under Contract No. DE-AC02-05CH11231.
	\end{acknowledgments}

\clearpage

\bibliography{bibtex_references} % Produces the bibliography via BibTeX.

\end{document}